\documentclass[prd
,preprint
,tightenlines
,showkeys
,nofootinbib 
,amsfonts,amsmath,amssymb 
]{revtex4}
\usepackage[pdftex]{graphicx}

\usepackage{bm,framed}
\usepackage{float,color}
\usepackage{physics}
 
\newcommand{\al}{\alpha}
\newcommand{\pa}{\partial}
\newcommand{\del}{\delta}
\newcommand{\bs}[1]{\boldsymbol{#1}}
\newcommand{\bu}{\bar{u}}
\newcommand{\anno}[1]{{#1}}

\begin{document}
\title{Tripartite Entanglement of Hawking Radiation in Dispersive Model}
\author{Yasusada Nambu}
\email{nambu@gravity.phys.nagoya-u.ac.jp}
\author{Yuki Osawa}
\email{osawa.yuki@h.mbox.nagoya-u.ac.jp}

\affiliation{Department of Physics, Graduate School of Science, Nagoya
University, Chikusa, Nagoya 464-8602, Japan}
%
%

%
%
\date{June 7, 2021} 

\begin{abstract} 
  We investigate entanglement of the Hawking radiation in a dispersive
  model with subluminal dispersion. In this model, feature of the
  Hawking radiation is represented by three mode Bogoliubov
  transformation connecting the in-vacuum state and the out-state. We
  obtain the exact form of the tripartite in-vacuum state which
  encodes structure of multipartite entanglement. Bogoliubov
  coefficients are computed by numerical calculation of the wave
  equation with subluminal dispersion and it is found that genuine
  tripartite entanglement persists in whole frequency range up to the
  cutoff arisen from the subluminal dispersion. In the low frequency
  region, amount of the tripartite entanglement is far small compared
  to bipartite entanglement between the Hawking particle and its
  partner mode, and the deviation from the thermal spectrum is
  negligible. On the other hand, in the high frequency region near the
  cutoff, entanglement of the system is equally shared by two
    pairs of three modes, and the thermal nature of the Hawking
  radiation is lost. 
\end{abstract} 

\maketitle 

\tableofcontents 

\section{Introduction}

The black hole horizon causes an extremely large redshift on outgoing
waves propagating from the vicinity of the horizon to the
asymptotically flat region. This redshift results in the Planckian
distribution of quantum mechanically created Hawking radiation~\cite{Hawking1974,Hawking1975a}. The
original Hawking's scenario relies on the assumption of infinite
amount of modes whose wavelength is shorter than the Planck
scale. Thus there is a possibility that sub-Planckian physics alter
the  nature of the Hawking radiation (trans-Planckian problem).
Within a setup of analog models of black hole geometry using moving
media with high frequency cutoff, the issue of the trans-Planckian
problem has been investigated by many
researchers~\cite{Unruh1995a,Brout1995a,Corley1996a,Tanaka2000a,Jacobson2003,Unruh2005a,Macher2009,Macher2009b,Robertson2012,Leonhardt2012,Coutant2012}. The
main purpose of these works is to clarify the effect of high frequency
cutoff on sub-Plankian origin of the thermal radiation predicted by
the original Hawking's work. These investigations
show the robustness of the thermal nature of the Hawking
  radiation; deviation from the Planckian distribution of the Hawking
  radiation is small if the cutoff scale is much higher than the
  surface gravity scale which is determined by the gradient of the
  flow velocity at the sonic point.
        
The thermal nature of the Hawking radiation can be understood from a
viewpoint of quantum correlation between the Hawking particles
(Hawking radiation) and their entangled partners which fall into the
black hole \cite{Hotta2015}. To verify the analog Hawking radiation in
laboratory experiments, it is crucial to understand the entanglement
structure of quasi-particles, of which excitation is observable by
experiments to confirm the quantum nature of the Hawking effect.
Considering excitation of quasi-particles in dispersive media, due to
non-linear dispersion relation, new wave modes participate and they
are responsible for forming multipartite entanglement structure of the
Hawking radiation. Actually, even for $(1+1)$-dimensional models,
right moving modes and left moving modes can mix each other and the
Bogoliubov transformation between the in-modes and the out-modes
becomes transformation between three
modes~\cite{Macher2009,Macher2009b,Busch2014}. Behavior of
entanglement involving three modes is investigated previously by
\cite{Busch2014} for the purpose of distinguishing quantum signals of
the Hawking radiation from classical thermal noise in laboratory
experiments. Based on inequalities for correlations between each mode,
which are equivalent to the Peres-Horodecki separability
criterion~\cite{Peres1996,Horodecki1997,Simon2000}, they analyzed
separability of specified two modes.  The separability is related to
domination of stimulated emission due to the initial thermal state in
the frame of fluid over spontaneous one, and they investigated
parameter range of which the spontaneous emission is possible.

The three mode Bogoliubov transformation naturally leads to formation
of multipartite entanglement among involving modes. The purpose of
this paper is to clarify the structure of the multipartite
entanglement of the Hawking radiation in dispersive media. We obtain
the exact form of the in-vacuum state with three modes and examine
entanglement between modes including the Hawking radiation.
The plan of this paper is as follows. In Sec.~II, we shortly review
the Hawking radiation in dispersive models. In Sec.~III, structure of
the in-vacuum state involving three modes is clarified and
entanglement structure of this state is investigated. In Sec.~IV, we
numerically solve the wave equation with subluminal dispersion and
obtain the Bogoliubov coefficients which are required to determine the
entanglement structure of three mode state. Sec.~V is devoted to
summary and conclusion. We adopt units $c=\hbar=G=1$ throughout this
paper.

\section{Hawking radiation in dispersive model}
We shortly review the Hawking radiation in a moving media with
subluminal dispersion. This part is mainly based on
  articles~\cite{Corley1996a,Jacobson2003,Macher2009,Macher2009b,Robertson2012}. The wave propagation in the media with stationary
flow is governed by the following wave equation for scalar
fluctuations (e.g. phonon)
\begin{equation}
  (\pa_t+\pa_xV)(\pa_t+V\pa_x)\phi=-F^2(-i\pa_x)\phi,
  \label{eq:wave-eq0}
\end{equation}
where $V(x)$ denotes the background flow velocity and $F(-i\pa_x)$
represents nonlinear modification of dispersion relation. The wave
equation \eqref{eq:wave-eq0} is \anno{the specific case of the
  equations for sound waves in the perfect fluid obtained by assuming
  constant background density and constant sound velocity of the
  fluid. In general, the wave equation in fluid is not conformally
  invariant~\cite{Anderson2014,Anderson2015,Fabbri2016}; after
  separating time dependence, the wave equation can be written in the
  form of one dimensional Schr\"{o}dinger equation with non-zero
  effective potential which originated from spatially dependent
  density and sound velocity. The non-zero effective potential causes
  mixing of right moving waves and left moving waves hence grey-body
  factors are nonzero. For the specific wave equation
  \eqref{eq:wave-eq0} adopted in this paper, the conformal invariance
  is violated due to dispersion and mixing of right moving waves and
  left moving waves occurs. The conformal invariance is recovered in
  the low frequency limit which corresponds to the dispersionless
  limit.  As we will see, this mixing results in the tripartite
  entanglement of the Hawking radiation in dispersive models.}

\begin{figure}[H]
\centering
  \includegraphics[width=0.4\linewidth,clip]{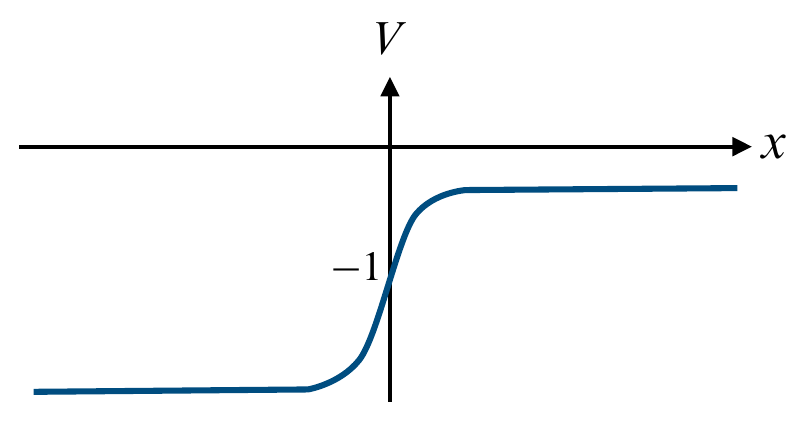}
  \caption{The profile of the background flow. The sonic horizon is
    located at $x=0$. For $x>0$, $|V|<1$
    (subsonic) and for $x<0$, $|V|>1$ (supersonic).}
  \label{fig:flow}
\end{figure}

\begin{figure}[H]
  \centering
  \includegraphics[width=1\linewidth,clip]{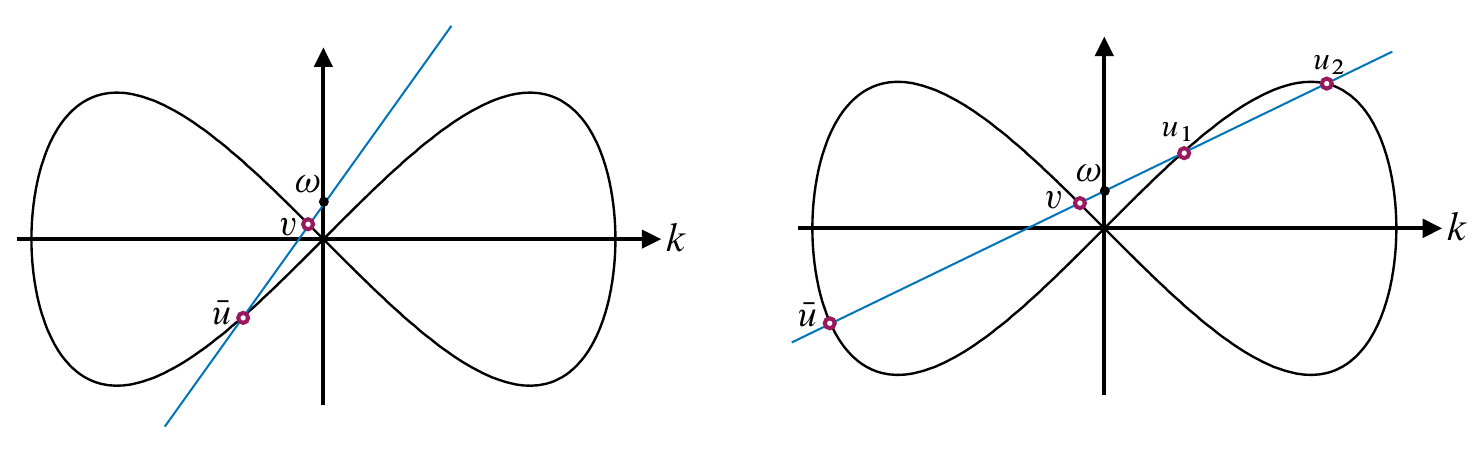}
  \caption{Modes appear in this model. Left panel: supersonic region ($V<-1$),
    right panel: subsonic region ($-1<V<0$). Only $u_1$ mode is right
    moving. $\bar u$ mode corresponds to  particles with negative
    energy (negative norm).}
  \label{fig:mode-graph}
\end{figure}

  The velocity profile is assumed to be
\begin{equation}
  V(x)=-1+D\tanh\left(\frac{\kappa x}{D}\right),\quad D<1.
\end{equation} 
The sonic horizon is located at $x=0$ for this flow (Fig.~\ref{fig:flow}). The
parameter $\kappa$ represents gradient of the flow velocity at the
sonic horizon. The asymptotic flow velocity is
$V_{+}:=V|_{x=+\infty}=-1+D, V_{-}:=V|_{x=-\infty}=-1-D$.
\noindent
As the flow is stationary, time dependence of the wave is separated as 
$\phi\propto e^{-i\omega t}$ and the wave equation becomes
\begin{equation}
  (-i\omega+\pa_xV)(-i\omega+V\pa_x)\phi=-F^2(-i\pa_x)\phi.
  \label{eq:wave-eq}
\end{equation}
The dispersion relation is obtained by substituting $\phi\propto e^{ikx}$:
\begin{equation}
  (\omega-V\,k)^2=F(k)^2,\quad F(k)=(k^2-k^4/k_0^2)^{1/2},
  \label{eq:dispersion}
\end{equation}
where subluminal type of dispersion is assumed.  The parameter $k_0$
determines the high frequency cutoff of the dispersion. It is possible
to identify wave modes as solutions of Eq.~\eqref{eq:dispersion}
(Fig.~\ref{fig:mode-graph}). In the subsonic region $x>0$, there are
two roots with negative wave numbers
$k_{\bar{u}}(\omega), k_v(\omega)$, and two roots with positive wave
numbers $k_{u_1}(\omega), k_{u_2}(\omega)$.  In the supersonic region
$x<0$, there are two modes with negative wave number
$k_{\bar u}(\omega), k_v(\omega)$. These four roots represent modes
appear in this model. We denote them as $u_1, u_2, \bar u,v$. In the
asymptotic region where $V(x)$ is constant, these modes behave as
plane waves
  \begin{equation}
    \varphi_m(x)\sim e^{ik_m(\omega)x},\quad m=u_1, u_2,\bar u, v.
  \end{equation}
  The group velocity of each mode is
  \begin{equation}
    V_g=V(x)\pm F'(k_m(\omega)).
  \end{equation}
  The mode $u_1$ has positive group velocity (right moving) and the
  modes $u_2, v, \bar u$ have negative group velocities (left moving).
\noindent
In the asymptotic subsonic region $x\rightarrow\infty $ where the flow
velocity becomes constant, we introduce the cutoff frequency given by
  \begin{equation}
    \omega_\text{cutoff}=\frac{k_0}{16}\left(3V_{+}+\sqrt{V_{+}^2+8}\right)\left(8-2V_{+}^2+2V_{+}\sqrt{V_{+}^2+8}\right)^{1/2}.
  \end{equation}
For $\omega>\omega_\text{cutoff}$, the dispersion relation does not
have real solutions with $k>0$ and there is no right propagating wave
modes in the asymptotic region of $x>0$.

\begin{figure}[H]
  \centering
  \includegraphics[width=0.35\linewidth,clip]{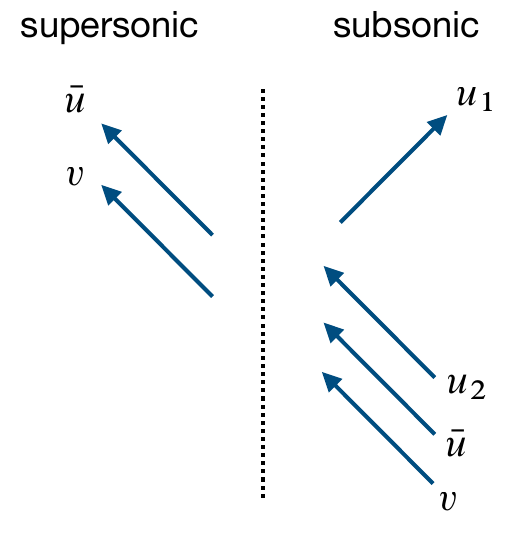}
  \hspace{1cm}
    \includegraphics[width=0.4\linewidth,clip]{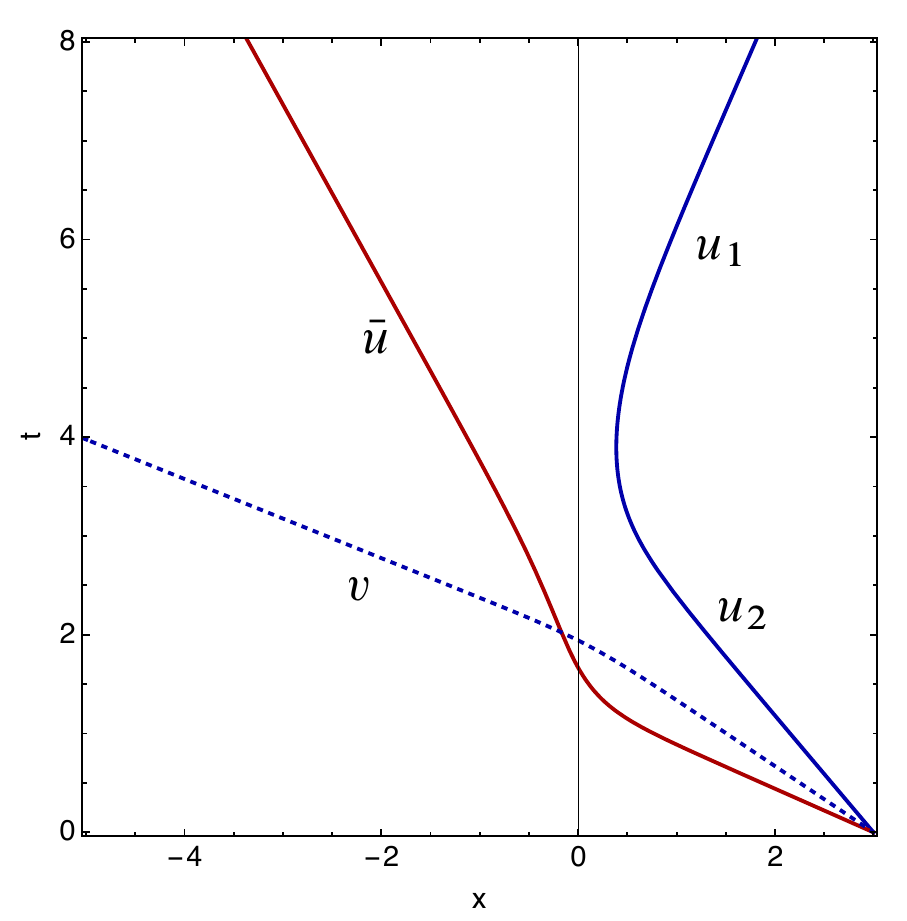}
    \caption{Left panel: schematic picture of modes in this model.  We
      have three modes for the in-state and three modes for the
      out-state. Right panel: trajectories of each mode obtained by
      the Hamiltonian $\omega(k,x)$. In this figure, adopted
      parameters are $D=1/2, \kappa=1, k_0=1, \omega=0.10$. The left
      moving $u_2$ mode is reflected to become the right moving $u_1$
      mode.}
      \label{fig:trajectory}
\end{figure}

Spacetime trajectories of each mode are obtained by equations of motion
derived from the Hamiltonian $\omega(k,x)$ (Fig.~\ref{fig:trajectory}):
\begin{equation}
  \frac{dx}{dt}=\frac{\pa\omega}{\pa k},\quad
  \frac{dk}{dt}=-\frac{\pa\omega}{\pa x},\quad\omega=V(x)\,k\pm F(k).
\end{equation}
In the vicinity of the sonic horizon, the group velocity of the left
moving mode $u_2$ becomes zero. At this point, the mode $u_2$ is
reflected and converted to become the right moving $u_1$ mode. The
location of the turning point depends on the cutoff frequency
$\omega_\text{cutoff}$. Under the assumption that $|V_{+}|\ll 1$ and
$\omega/\omega_\text{cutoff}\ll 1$, it is given by
\begin{equation}
  \kappa
  x\approx\frac{3}{2}\left(\frac{\omega}{2\omega_\text{cutoff}}\right)^{2/3},
  \label{eq:turning}
\end{equation}
and as the cutoff frequency
becomes larger, the turning point approaches the sonic horizon.

 Let us introduce normalized wave modes
  $\varphi_{u_{1,2}}(\omega)\propto e^{-i\omega t},
    \varphi_v(\omega)\propto e^{-i\omega t},
  \varphi_{\bar u}(-\omega)\propto e^{i\omega t}$ with $\omega>0$.
 They are solutions of \eqref{eq:wave-eq0}
  and are orthogonal and normalized with respect to the Klein-Gordon
  inner product
\begin{align}
  &(f,g):=i\int dx(f^*(\pa_t+V\pa_x)g-g(\pa_t+V\pa_x)f^*),\\
  &(\varphi_{u_{1,2}}(\omega_1),\varphi_{u_{1,2}}(\omega_2))=(\varphi_v(\omega_1),\varphi_v(\omega_2))
    =\del(\omega_1-\omega_2),\\
 &(\varphi_{\bar u}(-\omega_1),\varphi_{\bar u}(-\omega_2))=-\del(\omega_1-\omega_2).
\end{align}
 The mode $\bar u$ has negative energy and negative norm.  Now we consider
 the self-adjoint field operator $\hat\phi$ satisfying
 \eqref{eq:wave-eq0}. The
 creation and annihilation operators associated with the mode $f$ of
 the wave equation \eqref{eq:wave-eq0} is
 defined by
\begin{equation}
  \hat a(f)=(f,\hat\phi),\quad\hat a^\dag(f)=-(f^*,\hat\phi),
\end{equation}
and they are independent of choice of the time slice.  With the other
solution $g$ of the wave equation, they satisfy
\begin{equation}
  [\hat a(f),\hat a^\dag(g)]=(f,g),\quad [\hat a(f),\hat
  a(g)]=-(f,g^*),\quad [\hat a^\dag(f),\hat a^\dag(g)]=-(f^*,g).
\end{equation}
The Hawking radiation in this model is explained as follows. As the
in-state, we prepare the vacuum state,
\begin{equation}
  \hat a(\varphi_{u}^{\text{in}})\ket{0_\text{in}}=\hat
  a(\varphi_{\bar u}^{\text{in}}{}^*)\ket{0_\text{in}}=\hat a(\varphi_{v}^{\text{in}})\ket{0_\text{in}}=0.
\end{equation}
Then, using the Bogoliubov transformation \eqref{eq:Bogo-al}
  which connects the in-out creation and annihilation operators,
  number of the out-state particle in the subsonic region is
\begin{align}
  \bra{0_\text{in}}\hat a^\dag(\varphi^\text{out}_{u})\hat
  a(\varphi^\text{out}_{u})\ket{0_\text{in}}
  &=|\al_{\bu u}|^2\bra{0_\text{in}}\hat
    a(\varphi^{\text{in}}_{\bar u}{}^*)\hat
    a^\dag(\varphi^{\text{in}}_{\bar
    u}{}^*)\ket{0_\text{in}}
    \notag \\
  &=|\al_{\bu u}|^2\bra{0_\text{in}}[\hat a(\varphi^{\text{in}}_{\bar u}{}^*),\hat
    a^\dag(\varphi^{\text{in}}_{\bar u}{}^*)]\ket{0_\text{in}} \notag \\
  &=-|\al_{\bu u}|^2(\varphi^{\text{in}}_{\bar
    u},\varphi^{\text{in}}_{\bar u})>0.
\end{align}
Nonzero particle number implies outgoing radiation from the sonic
horizon. Previous works on the Hawking radiation in dispersive media
predict the thermal spectrum of emitted particles with the Hawking
temperature determined by the ``surface gravity'' $\kappa$, which
corresponds to the gradient of the background flow velocity at the sonic
horizon~\cite{Unruh2005a}.
\section{Structure of three mode in-vacuum state}
We investigate structure of the in-vacuum state involving three modes
which encode property of the Hawking radiation in dispersive models.

\subsection{Three mode Bogoliubov transformation}
Using the normalized modes
$\varphi_{u},\varphi_{\bar u}, \varphi_v$, the field
operator is expressed as the Fourier expansion with respect to
$\omega$ as
\begin{equation}
  \hat\phi=\int_0^\infty d\omega\left(\hat\phi(\omega)e^{-i\omega
      t}+\hat\phi^\dag(\omega)e^{i\omega t}\right),
\end{equation}
where the Fourier component of the field operator is expressed as
~\cite{Macher2009,Macher2009b,Busch2014}
\begin{align}
 \hat\phi(\omega)&=\hat a_u^\text{out}(\omega)\varphi_{u}^\text{out}(\omega)+\hat
  a_{\bu}^\text{out}{}^{\dag}(-\omega)\left(\varphi_{\bar 
    u}^\text{out}(-\omega)\right)^*+\hat
                   a_v^\text{out}(\omega)\,\varphi_v^\text{out}(\omega)
                   \notag \\
  &=\hat a_u^\text{in}(\omega)\varphi_{u}^\text{in}(\omega)+\hat
    a_{\bu}^\text{in}{}^{\dag}(-\omega)
    \left(\varphi_{\bar
    u}^\text{in}(-\omega)\right)^*+\hat a_v^\text{in}(\omega)\,\varphi_v^\text{in}(\omega). 
\end{align}
Creation and annihilation operators $\hat a_j^\text{out}, \hat
a_j^\text{out}{}^\dag$ are
associated with the out-state and $\hat a^\text{in}_j, \hat a^\text{in}_j{}^\dag$ are
associated with the in-state where indices are $j=\{u, \bu, v\}$. From now on, we omit the argument
$\omega$ of operators to simplify notations. Operators 
$\hat a_j^\text{in,out},\hat a_j^\text{in,out}{}^\dag$ are related
by the following three mode Bogoliubov transformation
\begin{equation}
  \begin{bmatrix}
    \hat a_u^\text{in}\\ \hat a_{\bu}^\text{in}{}^\dag\\ \hat a_v^\text{in}
  \end{bmatrix}
  =
  \begin{bmatrix}
    \al_{uu}&\al_{u\bu}&\al_{uv}\\
    \al_{\bu u}&\al_{\bu\bu}&\al_{\bu v}\\
    \al_{vu}&\al_{v\bu}&\al_{vv}
  \end{bmatrix}
  \begin{bmatrix}
    \hat a_u^\text{out}\\ \hat a_{\bu}^\text{out}{}^\dag\\ \hat a_v^\text{out}
  \end{bmatrix}.
  \label{eq:Bogo-al}
\end{equation}
If the mixing between $u,\bar u$ and $v$ is negligible
$\al_{uv}=\al_{\bu v}=\al_{vu}=\al_{v\bu}=0$, the transformation reduces to
the two mode transformation which appears in standard calculations of
the Hawking radiation. This mixing results in the gray-body
  factor in the power spectrum of the Hawking
  radiation. For
  $\hat{\mathcal{A}}^\text{in,out}:=(\hat a^\text{in,out}_u,\hat
  a_{\bu}^\dag{}^\text{in,out},\hat a_v^\text{in,out})^T$, commutation
  relations between creation and annihilation operators can be written
  as
  $[\hat{\mathcal{A}}_i^\text{in},
  \hat{\mathcal{A}}_j^\text{in}{}^\dag]=[\hat{\mathcal{A}}_i^\text{out},
  \hat{\mathcal{A}}_j^\text{out}{}^\dag]=\eta_{ij}$ where
  $\eta_{ij}:=\mathrm{diag}(1,-1,1)$, and the Bogoliubov coefficients
  obey the following skew-unitarity relation~\cite{Recati2009}:
  \begin{equation}
    \mathcal{S}\,\eta\,\mathcal{S}^\dag=\eta,\quad\mathcal{S}:=
    \begin{bmatrix}
      \al_{uu}&\al_{\bu u}&\al_{vu}\\
      \al_{u\bu}&\al_{\bu\bu}&\al_{v\bu}\\
      \al_{uv}&\al_{\bu v}&\al_{vv}
    \end{bmatrix},\quad \hat{\mathcal{A}}^\text{in}=\mathcal{S}^T\hat{\mathcal{A}}^\text{out}.
  \end{equation}
  This relation yields
\begin{align}
  &|\al_{11}|^2-|\al_{12}|^2+|\al_{13}|^2=1, \quad
    -|\al_{21}|^2+|\al_{22}|^2-|\al_{23}|^2=1, \notag\\
  &|\al_{31}|^2-|\al_{32}|^2+|\al_{33}|^2=1,\label{eq:bogo1}\\
  &\al_{11}\,\al_{21}^*-\al_{12}\,\al_{22}^*+\al_{13}\,\al_{23}^*=0,\quad
    -\al_{21}^*\,\al_{31}+\al_{22}^*\,\al_{32}-\al_{23}^*\,\al_{33}=0, \notag\\
 &\al_{11}\al_{31}{}^{\!\!\!*}-\al_{12}\al_{32}{}^{\!\!\!*}+\al_{13}\al_{33}{}^{\!\!\!*}=0,
    \label{eq:bogo2}\\
  &|\al_{11}|^2-|\al_{21}|^2+|\al_{31}|^2=1,\quad-|\al_{12}|^2+|\al_{22}|^2-|\al_{32}|^2=1,
  \notag\\
  &|\al_{13}|^2-|\al_{23}|^2+|\al_{33}|^2=1,  \label{eq:bogo3}\\
  &-\al_{11}{}^{\!\!\!*}\al_{12}+\al_{21}^*\al_{22}-\al_{31}{}^{\!\!\!*}\al_{32}=0,\quad
    \al_{12}\al_{13}{}^{\!\!\!*}-\al_{22}\al_{23}^*+\al_{32}\al_{33}{}^{\!\!\!*}=0,\notag
  \\
  &\al_{11}\al_{13}{}^{\!\!\!*}-\al_{21}^*\al_{23}+\al_{31}{}^{\!\!\!*}\al_{33}=0, \label{eq:bogo4}
\end{align}
where we adopted indices $\{1,2,3\}=\{u,\bu,v\}$.
These equations are used to check accuracy of our numerical solutions
of the wave equation.  From the representation of the field operator
\begin{equation}
  \hat\phi=\sum_j\hat{\mathcal{A}}^\text{in}_j\Phi^\text{in}_j=\sum_j\hat{\mathcal{A}}^\text{out}_j\Phi^\text{out}_j,\quad\Phi^{\text{in,out}}:=(\varphi^\text{in,out}_u,\varphi_{\bu}^*{}^\text{in,out},\varphi_v^\text{in,out})^T,
\end{equation}
the in-mode functions and the out-mode functions are connected by
$\Phi^\text{out}=\mathcal{S}\Phi^\text{in}$:
\begin{align}
  \begin{bmatrix}
    \varphi_{u}^\text{out}\\
    \varphi_{\bar u}^*{}^\text{out}\\
    \varphi_v^\text{out}
  \end{bmatrix}
  =
  \begin{bmatrix}
    \al_{uu}&\al_{\bu u}&\al_{vu}\\
    \al_{u\bu}&\al_{\bu\bu}&\al_{v\bu}\\
    \al_{uv}&\al_{\bu v}&\al_{vv}
  \end{bmatrix}
  \begin{bmatrix}
    \varphi_{u}^\text{in}\\
    \varphi_{\bar u}^*{}^\text{in}\\
    \varphi_v^\text{in}
  \end{bmatrix}.
    \label{eq:Bogo-phi}
\end{align}
\subsection{In-vacuum state}
\begin{figure}[H]
  \centering
  \includegraphics[width=0.5\linewidth,clip]{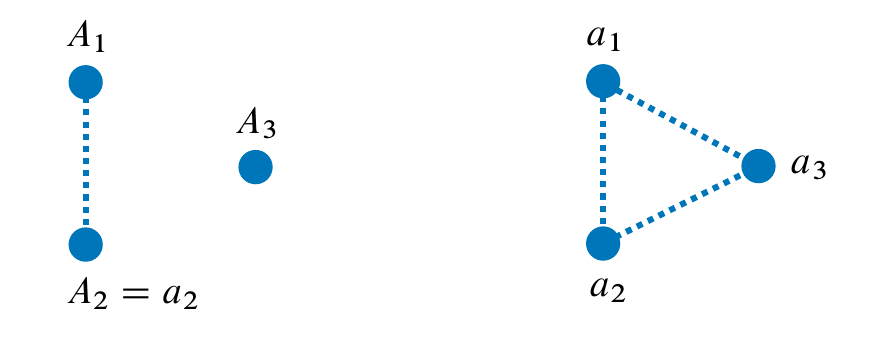}
  \caption{Three mode state defined by $A_i$ (left panel) and defined
    by $a_i^\text{out}$ (right panel). $A_1$ and $A_2$ constitute a pure two mode
    squeezed state and $A_3$ has no correlation with them. Expressing
    the state using the original modes defined by $a_i^\text{out}$, entanglement
    of the system is shared by all three modes. }
  \label{fig:three-mode}
\end{figure}

We adopt the following parametrization of $\al_{\bu u}, \al_{\bu\bu},
\al_{\bu v}$ which satisfies $|\al_{\bu\bu}|^2-|\al_{\bu
  u}|^2-|\al_{\bu v}|^2=1$:
\begin{equation}
  \al_{\bu u}=e^{i\varphi_1}\cos\theta\sinh
  r,\quad\al_{\bu\bu}=e^{i\varphi_2}\cosh
  r,\quad\al_{\bu v}=e^{i\varphi_3}\sin\theta\sinh r,\quad r\ge 0.
\end{equation}
From linear combinations of $\hat a_u^{\text{out}}, \hat a_v^{\text{out}}$, we introduce new
annihilation operators $\hat A_1, \hat A_3$ as
\begin{align}
  &\hat A_1=e^{-i(\varphi_1-\varphi_2)}\cos\theta\,\hat
  a_u^\text{out}+e^{-i(\varphi_3-\varphi_2)}\sin\theta\,\hat a_v^\text{out},\notag \\
 &\hat A_3=e^{i\varphi_3}\sin\theta\, \hat a_u^\text{out}-e^{i\varphi_1}\cos\theta
  \,\hat a_v^\text{out},\quad
  \hat A_2=\hat a_{\bu}^\text{out}.
\end{align}
They satisfy
$[\hat A_i, \hat A_j{}^{\!\dag}]=\del_{ij}, [\hat A_i, \hat
A_j]=0$. Inverting the relation,
\begin{align}
  &\hat a_u^\text{out}=e^{i(\varphi_1-\varphi_2)}\cos\theta\,\hat
  A_1+e^{-i\varphi_3}\sin\theta\,\hat A_3,\notag \\
  &\hat a_v^\text{out}=e^{i(\varphi_3-\varphi_2)}\sin\theta\,\hat
  A_1-e^{-i\varphi_1}\cos\theta\,\hat A_3,\quad\hat
  a_{\bu}^\text{out}=\hat A_2.
  \label{eq:mode-rel}
\end{align}
By expressing the Bogoliubov transformation \eqref{eq:Bogo-al} in
terms of new annihilation operators $\hat A_j$, we obtain
\begin{align}
  &\hat a_u^\text{in}=\frac{\al_{u\bu}}{\tanh r}\left(\hat A_1+\tanh r\hat
    A_2{}^{\!\!\dag}\right)
    +(\al_{uu}e^{-i\varphi_3}\sin\theta-\al_{uv}
    e^{-i\varphi_1}\cos\theta)\hat
    A_3,\\
    &\hat a_{\bu}^\text{in}=\al_{\bu\bu}^*\left(\tanh r\hat A_1{}^{\!\!\dag}+\hat A_2\right),\\
  &\hat a_v^\text{in}=\frac{\al_{v\bu}}{\tanh r}\left(\hat A_1+\tanh r\hat
    A_2{}^{\!\!\dag}\right)
    +(\al_{vu}e^{-i\varphi_3}\sin\theta-\al_{v\bu}e^{-i\varphi_1}\cos\theta)\hat
    A_3,
\end{align}
where we have used the relation
$\al_{11}\al_{21}^*-\al_{12}\al_{22}^*+\al_{13}\al_{23}^*=0$.  As the in-vacuum state
 $\ket{\psi_0}$ is determined by $\hat a_j^\text{in}\ket{\psi_0}=0, j=u,\bu,v$, we
 have equations determining the in-vacuum state
\begin{align}
  & \left(\tanh r\hat{A}_1{}^{\!\!\dag}+\hat
    A_2\right)\ket{\psi_0}=0,\quad
  \left(\hat{A}_1+\tanh r\hat
    A_2{}^{\!\!\dag}\right)\ket{\psi_0}=0,\quad
   \hat A_3\ket{\psi_0}=0.
\end{align}
Solving these equations, the form of the in-vacuum state
$\ket{\psi_0}$ is
\begin{equation}
  \label{eq:state}
  \ket{\psi_0}=\frac{1}{\cosh r}\sum_{n=0}^\infty (-\tanh
  r)^n\ket{n}_{A_1}\otimes\ket{n}_{A_2}\otimes\ket{0}_{A_3},
\end{equation}
where $\ket{n}_{A_j}$ are particle number states defined by the
out-state operators $\hat{A}_j$. Hence $\ket{\psi_0}$ is a produce
state of $\ket{0}_{A_3}$ and the two mode squeezed state defined by
$A_1$ and $A_2$ (Fig.~\ref{fig:three-mode}).
To obtain the wave function of the in-vacuum state involving three
modes, we introduce the following canonical quadrature operators
\begin{equation}
  \hat X_j=\frac{\hat A_j+\hat A_j{}^{\!\!\dag}}{\sqrt{2}},\quad
  \hat P_i=\frac{\hat A_j-\hat A_j{}^{\!\!\dag}}{i\sqrt{2}}.
\end{equation}
Then the wave function of the in-vacuum sate in $X$-representation is
\begin{align}
  \psi_0(X_1,X_2,X_3)&=\bra{X_1,X_2,X_3}\ket{\psi_0} \notag \\
  &=\frac{1}{\cosh r}\sum_{n=0}^\infty(-\tanh r)^n\bra{X_1}\ket{n}_{A_1}\bra{X_2}\ket{n}_{A_2}\bra{X_3}\ket{0}_{A_3},
\end{align}
where
\begin{equation}
  \bra{X_{1,2}}\ket{n}_{A_{1,2}}=\left(2^n
    n!\sqrt{\pi}\right)^{-1/2}H_n(X_{1,2})e^{-X_{1,2}^2/2},\quad
  \bra{X_3}\ket{0}_{A_3}=\left(\sqrt{\pi}\right)^{-1/2}e^{-X_3^2/2}.
\end{equation}
After all, the wave function is\footnote{We applied Mehler's
  formula
  $$\sum_{n=0}^\infty\frac{S^n}{2^nn!}H_n(X)H_n(Y)=(1-S^2)^{-1/2}\exp\left[\frac{2SXY-S^2(X^2+Y^2)}{1-S^2}\right]. $$}
\begin{align}
  \psi_0(X_1,X_2,X_3)&=\frac{\pi^{-3/4}}{\cosh
                     r}\,e^{-(X_1^2+X_2^2+X_3^2)/2}\sum_{n=0}^\infty\frac{(-\tanh
                     r)^n}{2^n n!}H_n(X_1)H_n(X_2) \notag \\
                   &=\pi^{-3/4}\exp\left[-\frac{\cosh 2r}{2}(X_1^2+X_2^2)
                     +\sinh 2r X_1X_2-\frac{X_3^2}{2}\right].
                     \label{eq:tri-state}
\end{align}
\subsection{Wigner function}
To analyze structure of the tripartite entanglement of the in-vacuum state,
it is convenient to adopt the phase space representation of the
wave function. For this purpose, we introduce the Wigner function of the
three mode  wave function \eqref{eq:tri-state} 
\begin{align}
  W(\bs{X},\bs{P}):&=\frac{1}{(2\pi)^3}\int
                    d^3\bs{y}\,e^{i\bs{P}\cdot\bs{y}}\,\psi_0\left(\bs{X}+\frac{\bs{y}}{2}\right)\psi_0^*\left(\bs{X}-\frac{\bs{y}}{2}\right)
                    \notag \\
  &=\frac{1}{\pi^3}\exp\Bigl[-\cosh 2r\,(X_1^2+P_1^2+X_2^2+P_2^2)+2\sinh
    2r\,(X_1X_2-P_1P_2)-(X_3^3+P_3^2)\Bigr].
\end{align}
In terms of  the phase space vector
$\bs{\xi}=(X_1,P_1,X_2,P_2,X_3,P_3)^T$, the Wigner function is
expressed as
\begin{equation}
  W(\bs{\xi})=\frac{1}{\pi^3}
  \exp\left[-\frac{1}{2}\bs{\xi}^T\bs{V}^{-1}\bs{\xi}\right],
  \qquad \int d^6\bs{\xi}W(\bs{\xi})=1,
\end{equation}
where $\bs{V}$ is the covariance matrix
\begin{equation}
  V_{ij}=\frac{1}{2}\expval{\hat\xi_i\hat\xi_j+\hat\xi_j\hat\xi_i}=\int
  d^6\bs{\xi}\,\xi_i\xi_jW(\bs{\xi}),\quad i,j=1,\cdots 6,
\end{equation}
and its components are
\begin{equation}
  \bs{V}=
  \frac{1}{2}\begin{bmatrix}
    \cosh 2r \bs{I} &-\sinh 2r\bs{Z}& 0\\
    -\sinh 2r \bs{Z} &\cosh 2r\bs{I}&0\\
    0&0&\bs{I}
    \end{bmatrix},\quad
    \bs{I}=\begin{bmatrix}1&0\\0&1\end{bmatrix},\quad
    \bs{Z}=\begin{bmatrix}1&0\\0&-1\end{bmatrix}.
    \label{eq:V}
\end{equation}
The original canonical quadrature operators associated with
$\hat a_j^\text{out}, \hat a_j^\text{out}{}^\dag$ are
\begin{equation}
      \hat x_j=\frac{\hat a_j^\text{out}+\hat a_j^\text{out}{}^{\dag}}{\sqrt{2}},\quad
  \hat p_j=\frac{\hat a_j^\text{out}-\hat a_j^\text{out}{}^{\dag}}{i\sqrt{2}}.
\end{equation}
The relation between $\bs{\xi}_0=(x_1,p_1,x_2,p_2,x_3,p_3)^T$ and
$\bs{\xi}=(X_1,P_1,X_2,P_2,X_3,P_3)^T$ is obtained from
\eqref{eq:mode-rel}:
\begin{equation}
  \begin{bmatrix}X_1\\P_1\\X_2\\P_2\\X_3\\P_3\end{bmatrix}=
  \begin{bmatrix}
    \cos\theta\cos\varphi_{12}&\cos\theta\sin\varphi_{12}&0&0&\sin\theta\cos\varphi_{32}
    &\sin\theta\sin\varphi_{32}\\
    -\cos\theta\sin\varphi_{12}&\cos\theta\cos\varphi_{12}&0&0&-\sin\theta\sin\varphi_{32}&\sin\theta\cos\varphi_{32}\\
    0&0&1&0&0&0\\
    0&0&0&1&0&0\\
    \sin\theta\cos\varphi_3&-\sin\theta\sin\varphi_3&0&0&-\cos\theta\cos\varphi_1&\cos\theta\sin\varphi_1\\
    \sin\theta\sin\varphi_3&\sin\theta\cos\varphi_3&0&0&-\cos\theta\sin\varphi_1&-\cos\theta\cos\varphi_1
  \end{bmatrix}
  \begin{bmatrix}x_1\\p_1\\x_2\\p_2\\x_3\\p_3\end{bmatrix},
\end{equation}
where $\varphi_{ij}:=\varphi_i-\varphi_j$.  We denote this relation as
$\bs{\xi}=\bs{S}\bs{\xi}_0$.  As $\bs{S}$ is \anno{a} symplectic
transformation, it satisfies the symplectic condition
\begin{equation}
  \bs{S}\bs{\Omega}_3\bs{S}^T=\bs{\Omega}_3,\quad
  \bs{\Omega}_3:=\bigoplus_{j=1}^3\bs{J},\quad
   \bs{J}=
  \begin{bmatrix}0&1\\-1&0\end{bmatrix}.
\end{equation}
The Wigner function $W_0$ for the original variables $\bs{\xi}_0$ is 
\begin{equation}
 W_0(\bs{\xi}_0)= W(\bs{S}\bs{\xi}_0)=\frac{1}{\pi^3}
  \exp\left[-\frac{1}{2}\bs{\xi}_0^T\bs{S}^T\bs{V}^{-1}\bs{S}\bs{\xi}_0\right]=
  \frac{1}{\pi^3}\exp\left[-\frac{1}{2}\bs{\xi}_0^T\bs{V}_0^{-1}\bs{\xi}_0\right],
\end{equation}
where the covariance matrix $\bs{V}_0$ is related to $\bs{V}$ as
\begin{equation}
  \bs{V}_0=\bs{S}^{-1}\bs{V}(\bs{S}^T)^{-1}=\bs{\Omega}_3\bs{S}^T\bs{\Omega}_3\bs{V}\bs{\Omega}_3\bs{S}\bs{\Omega}_3,
\end{equation}
and its components are given by
\begin{equation}
  \bs{V}_0= \begin{bmatrix}
    \bs{V}_1&\bs{W}_{12}&\bs{W}_{31}\\
    \bs{W}_{12}^T&\bs{V}_2&\bs{W}_{23}\\
    \bs{W}_{31}^T&\bs{W}_{23}^T&\bs{V}_3
    \label{eq:V0}
\end{bmatrix}
\end{equation}
with $2\times2$ submatrices 
\begin{align}
&\bs{V}_1=\left(\frac{1}{2}+\sinh^2
                r\cos^2\theta\right)\,\bs{I},\quad
                \bs{V}_2=\frac{\cosh 2r}{2}\,\bs{I},\quad
                \bs{V}_3=\left(\frac{1}{2}+\sinh^2r \sin^2\theta\right)\,\bs{I},\\
&\bs{W}_{12}=-\frac{\sinh 2r \cos\theta}{2}
          \begin{bmatrix}
            \cos\varphi_{12}&\sin\varphi_{12}\\\sin\varphi_{12}&-\cos\varphi_{12} 
          \end{bmatrix},\quad
     \bs{W}_{23}=-\frac{\sinh 2r
                 \sin\theta}{2}\begin{bmatrix}\cos\varphi_{32}&
                                 \sin\varphi_{32}\\
\sin\varphi_{32}&-\cos\varphi_{32}   \end{bmatrix} , \notag\\             
       &\bs{W}_{31}=\frac{\sinh^2r\sin2\theta}{2}\begin{bmatrix}
                     \cos\varphi_{13}&-\sin\varphi_{13}\\
\sin\varphi_{13}&\cos\varphi_{13} \end{bmatrix} .            
\end{align}
As the total system is pure, the covariance matrix $\bs{V}_0$
satisfies the following purity condition~\cite{Giedke2014}
\footnote{Symplectic eigenvalues of $\bs{V}_0$ are obtained as
  positive eigenvalues of $i\bs{\Omega}_3\bs{V}_0$. As symplectic
  eigenvalues of a pure state is $1/2$, $i\bs{\Omega}_3\bs{V}_0$ is
  diagonalized using a unitary matrix $U$, the spectrum of eigenvalue
  is
  $U(i\bs{\Omega}_3\bs{V}_0)U^\dag=1/2\,
  \mathrm{diag}(1,1,1,-1,-1,-1)$.  Thus
  $U(i\bs{\Omega}_3\bs{V}_0)^2U^\dag=\bs{I}_{6\times 6}/4$. From this,
  the relation \eqref{eq:pure-V} is derived.}
\begin{equation}
  \bs{V}_0\bs{\Omega}_3\bs{V}_0=\frac{1}{4}\,\bs{\Omega}_3.
  \label{eq:pure-V}
\end{equation}
Submatrices $\bs{W}_{12}, \bs{W}_{23}, \bs{W}_{31}$ can be
simultaneously diagonalized using the local rotation with respect to
each mode \cite{Giedke2014}. As we are interested in  entanglement of
the state, and it can be quantified in terms of the
symplectic eigenvalues and the negativity determined by the symplectic
eigenvalues. They are independent of
$\varphi_1-\varphi_2,\varphi_3-\varphi_2, \varphi_1-\varphi_3$ and 
for the purpose of obtaining negativities of the state defined by the
covariance matrix \eqref{eq:V0} , we can set
$\varphi_1=\varphi_2=\varphi_3$ without loss of generality. After all,
entanglement structure of the present three mode model can be
investigated using the following covariance matrix
\begin{equation}
  \bs{V}_0= \frac{1}{2}\begin{bmatrix}
    \left(1+2\sinh^2r\cos^2\theta\right)\bs{I}&-\sinh
    2r\cos\theta\bs{Z}&\sinh^2r\sin 2\theta\bs{I} \\
    *&\cosh 2r\bs{I}&-\sinh 2r\sin\theta\bs{Z}\\
    *&*&\left(1+2\sinh^2r\sin^2\theta\right)\bs{I}
  \end{bmatrix}.
  \label{eq:V0}
\end{equation}
Two parameters $r,\theta$ in the covariance matrix are related to
the Bogoliubov coefficients as 
\begin{equation}
  \cosh r=|\al_{\bu\bu}|,\quad
  \tan\theta=\left|\frac{\al_{\bu v}}{\al_{\bu u}}\right|.
\end{equation}
For $\theta=0,\pi$, $\bs{V}_0$ reduces to $\bs{V}$ (Eq.~\eqref{eq:V}).
The parameter $\theta$ represents degree of $\bu v$-mixing.

By tracing out modes 23, 31, 12, reduced single mode states are
\begin{equation}
  \hat\rho_1:=\mathrm{Tr}_{2,3}\hat\rho,\quad
  \hat\rho_2:=\mathrm{Tr}_{3,1}\hat\rho,\quad
  \hat\rho_3=\mathrm{Tr}_{1,2}\hat\rho,
\end{equation}
and their covariance matrices are given by $\bs{V}_1, \bs{V}_2,
\bs{V}_3$, respectively. By tracing out modes 3,1,2, reduced bipartite
states are
\begin{equation}
  \hat\rho_{12}:=\mathrm{Tr}_3\hat\rho,\quad
  \hat\rho_{23}:=\mathrm{Tr}_1\hat\rho,\quad
  \hat\rho_{31}:=\mathrm{Tr}_2\hat\rho,
\end{equation}
and their covariance matrices are
\begin{align}
&\bs{V}_{12}=\frac{1}{2}\begin{bmatrix}
(1+2\sinh^2r\cos^2\theta)\bs{I}&-\sinh 2r\cos\theta\bs{Z} \\-\sinh
2r\cos\theta\bs{Z}&\cosh 2r\bs{I}
\end{bmatrix} ,\\
&\bs{V}_{23}=\frac{1}{2}\begin{bmatrix}
\cosh 2r\bs{I}&-\sinh 2r\sin\theta\bs{Z}\\-\sinh 2r\sin\theta\bs{Z}&(1+2\sinh^2r\sin^2\theta)\bs{I}
\end{bmatrix},  \\
&\bs{V}_{31}=\frac{1}{2}\begin{bmatrix}
  (1+2\sinh^2r\sin^2\theta)\bs{I}&\sinh^2r\sin2\theta\bs{I}\\
  \sinh^2r\sin2\theta\bs{I}&(1+2\sinh^2r\cos^2\theta)\bs{I}
\end{bmatrix}.
\end{align}

\subsection{Negativity and tripartite entanglement}
In this paper, we adopt the negativity as an entanglement
  measure. For a Gaussian state with canonical variables
  $\hat{\bs{\xi}}=(\hat x_1,\hat p_1,\cdots, \hat x_N, \hat p_N)^T, \langle\hat{\bs{\xi}}\rangle=0$, the commutation relation is
  \begin{equation}
    [\hat\xi_j,
    \hat\xi_k]=i(\Omega_N)_{jk},\quad\bs{\Omega}_N=\bigoplus_{j=1}^N\bs{J}.
  \end{equation}
  The Gaussian state is completely characterized by the covariance matrix 
 $V_{jk}=\langle{\hat\xi_j\hat\xi_k+\hat\xi_k\hat\xi_j}\rangle/2$. 
For a physical state, the density matrix must be non-negative and the
corresponding covariance matrix must satisfy the inequality
\begin{equation}
  \bs{V}+\frac{i}{2}\bs{\Omega}_N\ge 0, \label{eq:phys-ineq}
\end{equation}
which is the generalization of the uncertainty relation between two
canonically conjugate variables. The separability of the bipartite
Gaussian state composed of parties A and B is expressed in terms of
the partial transpose operation defined by reversing the sign of one
party's momentum~\cite{Peres1996,Horodecki1997,Simon2000}. For the
partially transposed covariance matrix $\tilde{\bs{V}}$, the
sufficient condition of the separability is given by
\begin{equation}
  \tilde{\bs{V}}+\frac{i}{2}\bs{\Omega}_N\ge 0. \label{eq:unphys-ineq}
\end{equation}
If this inequality is violated, the bipartite state is
entangled. To quantify entanglement, we introduce symplectic
  eigenvalues $\nu_i$ of $\bs{V}$, which are obtained by diagonalizing
  the covariance matrix with a symplectic transformation. Practically,
  symplectic eigenvalues can be obtained as positive eigenvalues of the
  matrix $i\bs{\Omega}_N\bs{V}$~\cite{Weedbrook2012}. In terms of
  symplectic eigenvalues, the physical condition \eqref{eq:phys-ineq}
is $\nu_i\ge 1/2$ and the separability condition is
$\tilde\nu_i\ge 1/2$ where $\tilde\nu_i$ are symplectic eigenvalues of
$\tilde{\bs{V}}$.  The entanglement negativity~\cite{Vidal2002a} is
given by
\begin{equation}
  \mathcal{N}:=\frac{1}{2}\mathrm{max}\left[\left(\prod_{\tilde\nu_i<1/2}\frac{1}{2\tilde\nu_i}\right)-1,0\right].
\end{equation}
Nonzero value of the negativity implies non-separability of the
bipartite state and existence of bipartite entanglement of the system.
Related to the negativity, the logarithmic
negativity is defined by
\begin{equation}
  E_N:=\log(2\mathcal{N}+1).
\end{equation}

\begin{figure}[H]
  \centering
  \includegraphics[width=1\linewidth,clip]{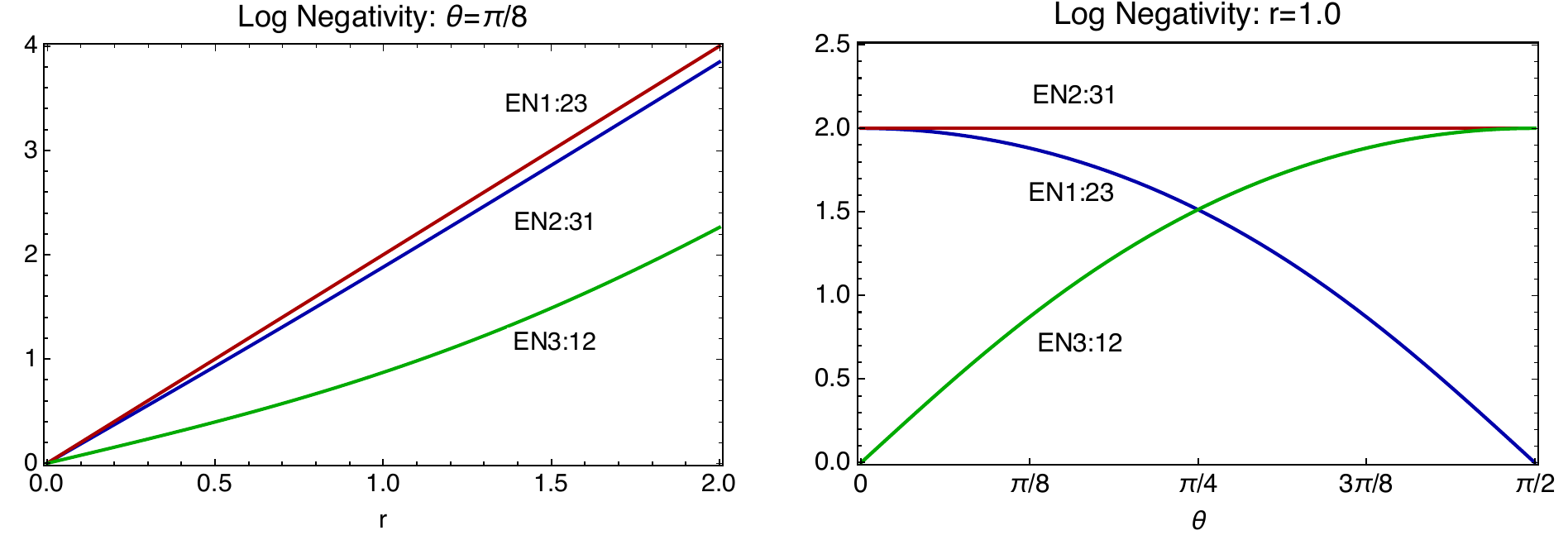}
  \caption{Bipartite entanglement of the three mode state. Amount of
    entanglement increases as the parameter $r$ increases (left
    panel). Relative magnitude of
    $E_N{}_{1:23}$ and $E_N{}_{3:12}$ changes depending on values of 
    the parameter $\theta$.}
    \label{fig:neg3}
\end{figure}

Bipartite entanglement of three mode state is evaluated by specifying
three possible bipartitions (Fig.~\ref{fig:part1}):
\begin{figure}[H]
  \centering
  \includegraphics[width=0.7\linewidth,clip]{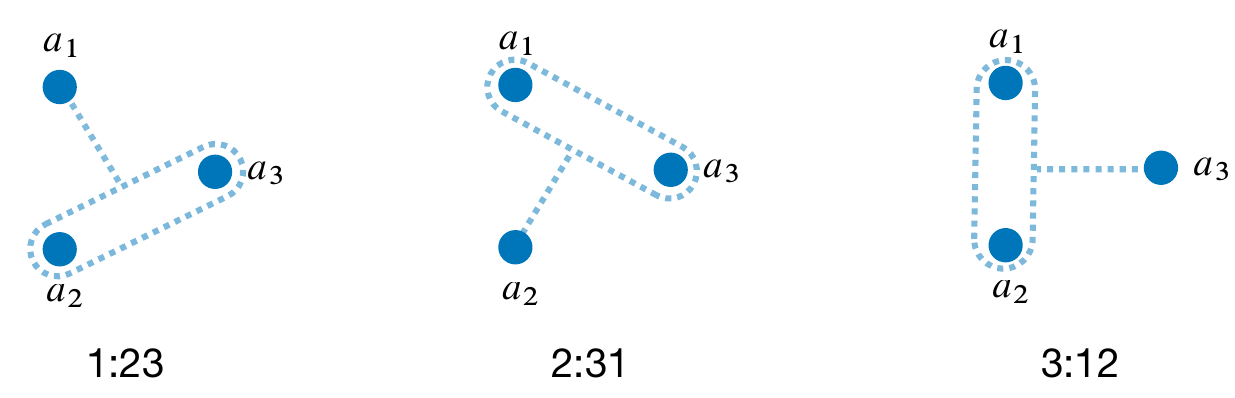}
  \caption{Possible bipartitions of three mode state. For all three
    different partitions, bipartite entanglement (negativity) is
    nonzero in the present model. This implies existence of genuine
    tripartite entanglement which cannot be reducible to bipartite
    entanglement between reduced two modes.}
  \label{fig:part1}
\end{figure}  
\noindent
Behavior of logarithmic negativities
$E_N{}_{1:23}, E_N{}_{2:31}, E_N{}_{3:12}$ is shown
in Fig.~\ref{fig:neg3}. For three different partitions of three modes,
the logarithmic negativity is nonzero and grows as $r$ increases with $\theta$
fixed. This behavior implies the three mode state has genuine
tripartite entanglement which cannot be reducible to bipartite
entanglement between two modes. For a fixed value of $r$, the following
inequalities hold (right panel of Fig.~\ref{fig:neg3})
\begin{equation}
  E_N{}_{2:31}\geq E_N{}_{1:23},\quad
  E_N{}_{2:31}\geq E_N{}_{3:12}.
\end{equation}
Magnitude relation between $E_N{}_{1:23}$
and $E_N{}_{3:12}$ changes depending on values of $\theta$.

\begin{figure}[H]
  \centering
  \includegraphics[width=0.7\linewidth,clip]{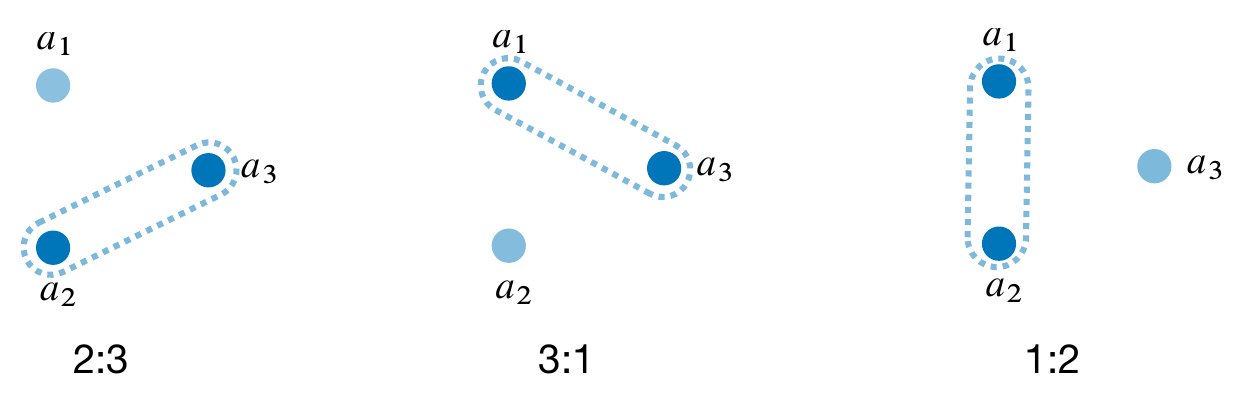}
  \caption{Reduced two modes states are obtained
    by tracing out one mode. In the present three mode model,
    $E_N{}_{3:1}=0$ and outgoing mode 1 ($u$) and ingoing mode
    3 ($v$) are separable as reduced two mode state. However, this
    does not imply the mode 1 and the mode 3 do not share entanglement.}
  \label{fig:part2}
\end{figure}

\begin{figure}[H]
  \centering
  \includegraphics[width=1\linewidth,clip]{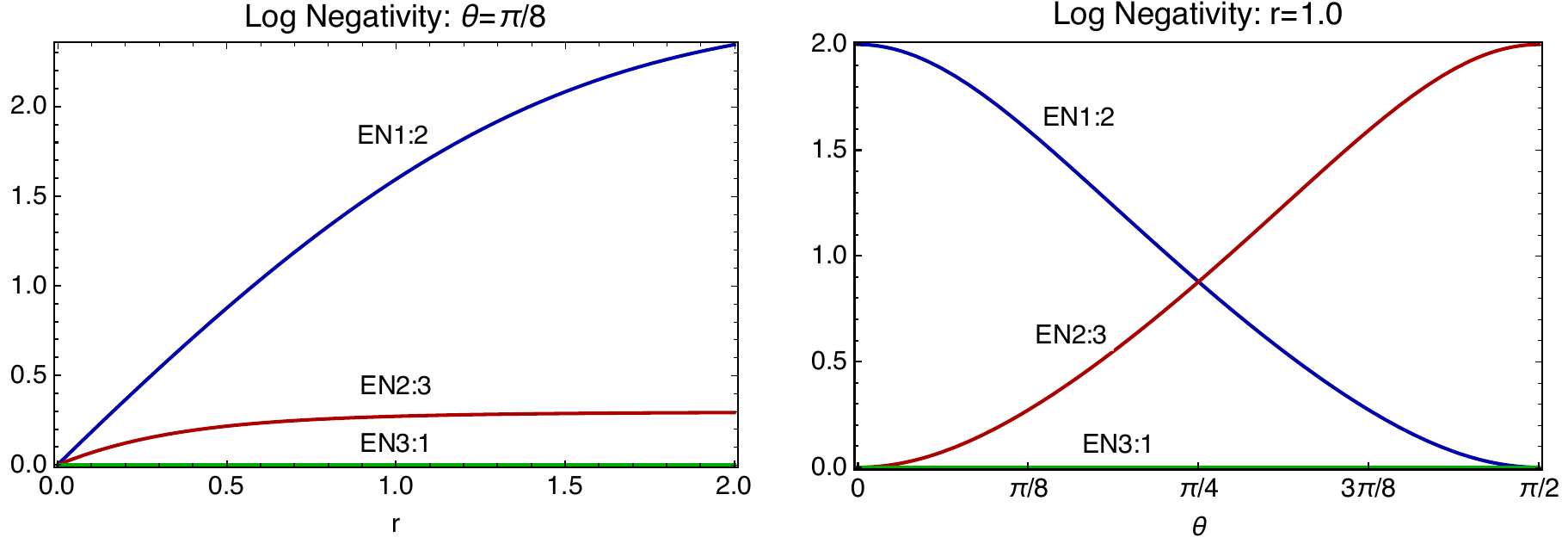}
  \caption{Bipartite entanglement of reduced two modes
    states. Negativity between the mode 1 and 3
    is$E_N{}_{3:1}=0$. Relative magnitude of $E_N{}_{1:2}$ and
    $E_N{}_{2:3}$ changes depending on values of the parameter
    $\theta$. }
  \label{fig:neg2}
\end{figure}

\begin{figure}[H]
  \centering
  \includegraphics[width=1\linewidth,clip]{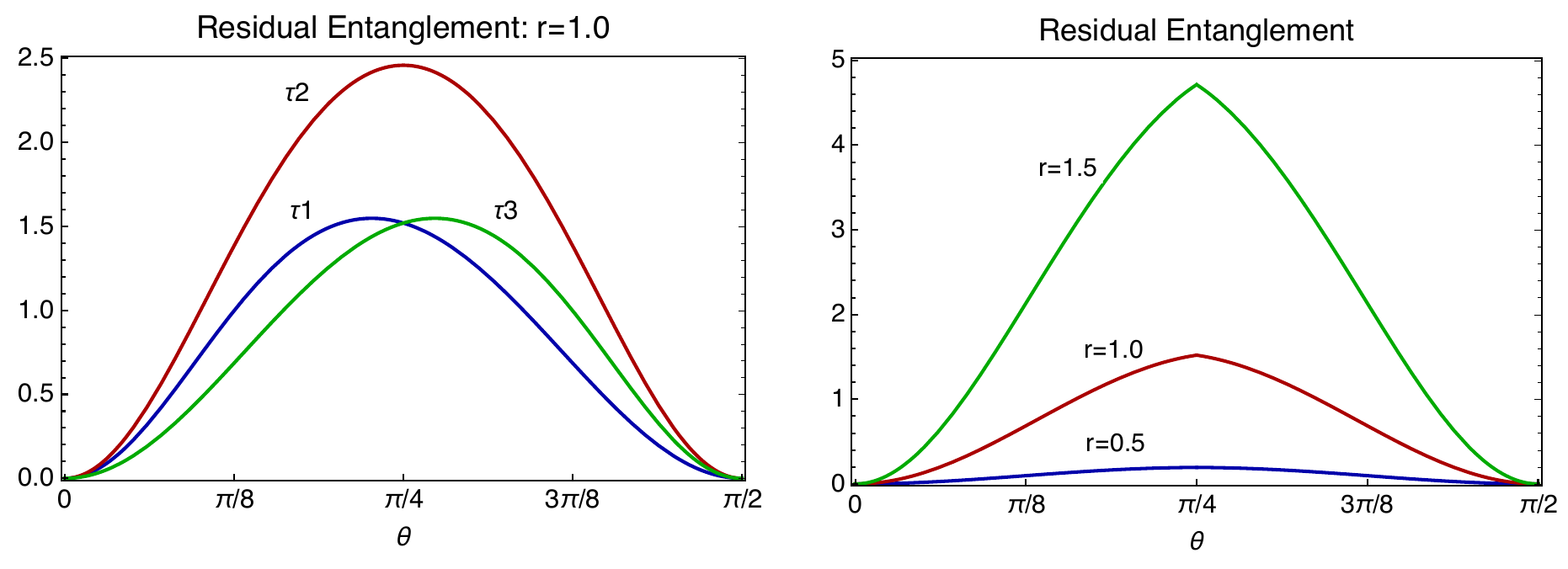}
  \caption{Residual of entanglement $\tau_1,\tau_2,\tau_3$ (left
    panel) and $\tau$ (right panel). Genuine tripartite entanglement exists for
  $\theta\neq 0,\pi/2$. For $\theta=0,\pi/2$, one mode decouples from
  other two modes and residual entanglement becomes zero.}
  \label{fig:res} 
\end{figure}

By tracing out an one mode of three mode state, entanglement between
reduced two modes is also evaluated (Fig~\ref{fig:part2}). 
\noindent
For reduced two mode states, behavior of the logarithmic negativity is
shown in Fig.~\ref{fig:neg2}.  We have always
\begin{equation}
  E_N{}_{3:1}=0,
\end{equation}
and the mode 1 and the mode 3 are separable as a reduced two mode system.  For
$\theta=0$, $E_N{}_{2:3}=0$ and the modes 1 and 2 forms
pure two mode squeezed state and the mode 3 decouples. For
$\theta=\pi/2$, $E_N{}_{1:2}=0$ and the mode 2 and 3 forms
pure two mode squeezed state and the mode 1 decouples.
To quantify genuine tripartite entanglement which cannot be
reducible to bipartite entanglement between two modes, we examine the 
residual of entanglement defined by 
\begin{align}
  &\tau_1=E_{1:23}-E_{1:2}-E_{1:3}=
    E_{1:23}-E_{1:2},\notag \\
  &\tau_2=E_{2:31}-E_{1:2}-E_{2:3},\\
  &\tau_3=E_{3:12}-E_{3:1}-E_{3:2}=E_{3:12}-E_{2:3}, \notag
\end{align}
where $E$ denotes square of the logarithmic negativity.  The residual
of entanglement quantifies genuine multipartite entanglement and is
proved to have positive values for three mode pure Gaussian state
\cite{Adesso2006b} (monogamy relation of entanglement). If the
residual is zero, entanglement in three mode state can be represented
as summation of entanglement of the reduced two modes state and there is
no genuine multipartite entanglement. The tripartite entanglement of
the system is also quantified by a single quantity
\begin{equation}
  \tau=\mathrm{min}(\tau_1,\tau_2,\tau_3).
\end{equation}
Dependence of $\theta$ of the
residual entanglement is shown in Fig.~\ref{fig:res}.
\noindent 
For $\theta=0$, the logarithmic negativity is
$E_N{}_{3:12}=E_N{}_{2:3}=0, E_N{}_{1:23}=E_N{}_{2:31}=E_N{}_{1:2}\neq
0$ and $\tau_{1,2,3}=0$. The modes 1,2 constitute a pure entangled state
and the mode 3 decouples.  Standard scenario of the Hawking radiation
corresponds to $\theta=0$ because the mode 3 (left moving $v$ mode)
decouples and the mode 2 ($\bar u$ mode) becomes the entanglement partner
of the mode 1 (the Hawking particle $u$).  For $\theta \neq 0, \pi/2$,
entanglement of three mode states is shared by all three modes
and there is genuine tripartite entanglement.

\section{Entanglement of Hawking radiation in dispersive model}
So far we obtained the structure of the in-vacuum state defined by
three mode Bogoliubov transformation \eqref{eq:Bogo-al} without
specifying behavior of coefficients. By obtaining wave modes in
dispersive media numerically, it is possible to examine tripartite
entanglement of the Hawking radiation.
\subsection{Numerical method}
\begin{figure}[H]
    \centering
    \includegraphics[width=0.9\linewidth,clip]{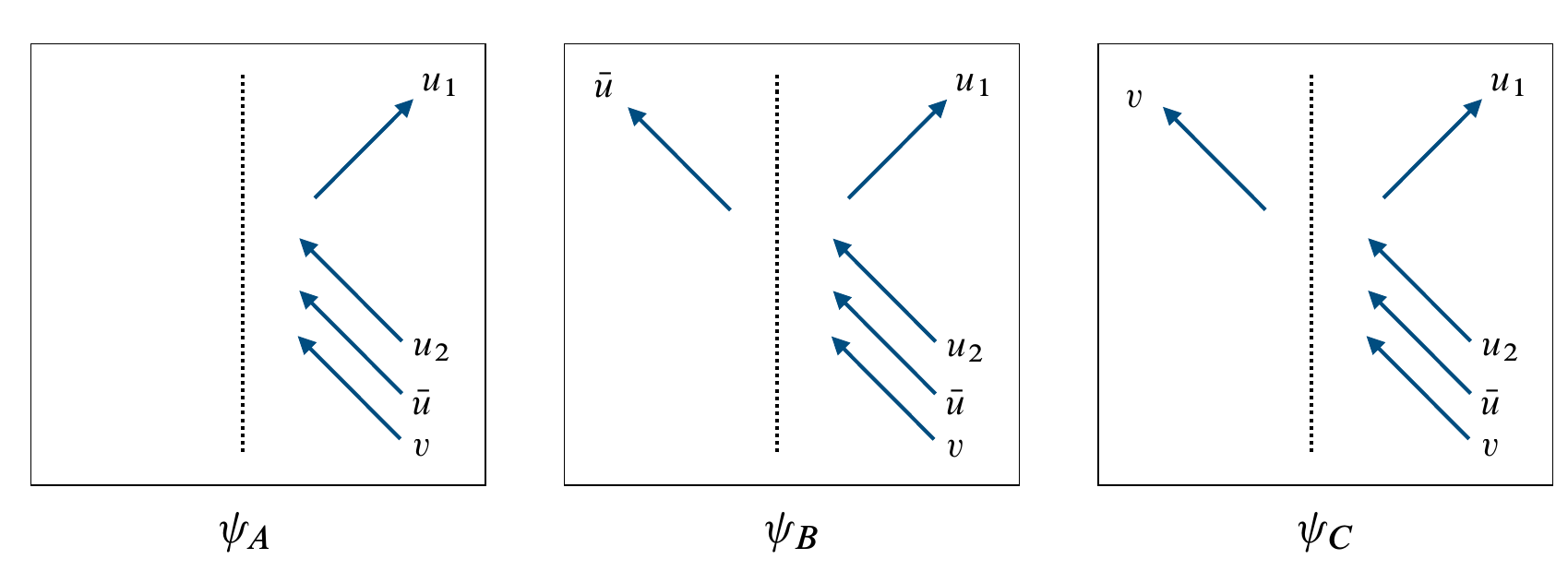}
    \caption{Three different combinations of modes to determine the Bogoliubov
      coefficients. The dotted line represents the sonic horizon.}
    \label{fig:modes}
  \end{figure}

  We solve the wave equation \eqref{eq:wave-eq} to obtain the
  Bogoliubov coefficients. For this purpose, we consider three
  different boundary conditions for the wave equation
  (Fig.~\ref{fig:modes}). Corresponding to the in-state, ingoing modes
  ($u_2$, $\bar u$, $v$) are assumed. Three different out-states are
  possible by choosing different boundary conditions at $x=-\infty$.
\noindent
The mode $\psi_A$ is defined by imposing a boundary condition that all
wave modes decay at $x\rightarrow -\infty$. This combination of modes
corresponds to $\varphi_{u}^\text{out}$ in \eqref{eq:Bogo-phi}.  The
mode $\psi_B$ is defined by imposing a boundary condition that only
$\bar u$ mode exists at $x\rightarrow -\infty$ and is represented by a
linear combination of $\varphi_{\bar u}^\text{out}$ and
$\varphi_v^\text{out}$:
$\psi_B\propto (\varphi_{\bar
  u}^\text{out})^*+c_1\varphi_{u}^\text{out}$. The mode $\psi_C$ is
defined by imposing a boundary condition that only $v$ modes exists at
$x\rightarrow -\infty$ and is represented by a linear combination of
$\varphi_v^\text{out}$ and $\varphi_{u}^\text{out}$:
$\psi_C\propto \varphi_v^\text{out}+c_2\varphi_{u}^\text{out}$. The
coefficients $c_1,c_2$ are determined by required boundary
conditions. Taking into account of the definition of the Bogoliubov
coefficients \eqref{eq:Bogo-phi}, behaviors of these modes at
$x=\pm\infty$ are
\begin{align}
  &\psi_A\propto\varphi_{u}^\text{out} \notag \\
  &\sim
  \begin{cases}
    0\quad& (x\rightarrow-\infty)\\
    \varphi_{u_1}+\al_{uu}\varphi_{u_2}+\al_{\bu u}\varphi_{\bar
      u}^*+\al_{vu}\varphi_v,\quad& (x\rightarrow+\infty)
  \end{cases}\\
  &\psi_B\propto(\varphi_{\bar
    u}^\text{out})^*+c_1\varphi_{u}^\text{out} \notag\\
  &\sim
  \begin{cases}
    \varphi_{\bar u}^*\quad& (x\rightarrow-\infty)\\
    (\al_{u\bu}+c_1\al_{uu})\varphi_{u_2}+(\al_{\bu\bu}+c_1\al_{\bu
      u})\varphi_{\bar
      u}^*+(\al_{v\bu}+c_1\al_{vu})\varphi_{v}+c_1\varphi_{u_1},\quad& (x\rightarrow+\infty)
  \end{cases}\\
  &\psi_C\propto\varphi_{v}^\text{out}+c_2\varphi_{u}^\text{out}\notag
  \\
  &\sim
  \begin{cases}
    \varphi_v\quad& (x\rightarrow-\infty)\\
    (\al_{uv}+c_2\al_{uu})\varphi_{u_2}+(\al_{\bu
        v}+c_2\al_{\bu u})\varphi_{\bar
    u}^*+(\al_{vv}+c_2\al_{vu})\varphi_{v}+c_2\varphi_{u_1}.\quad& (x\rightarrow+\infty)
  \end{cases}
\end{align}
Asymptotic behavior of these modes at $x\rightarrow +\infty$ can be
represented by the  Fourier expansion 
\begin{equation}
  \psi_{A,B,C}\sim \sum_m d_{A,B,C}^{(m)}\,\varphi_m,\quad
  m=u_1,u_2,\bar u,v,
\end{equation}
where $\varphi_m\propto e^{ik_m(\omega)x}$ are plane waves at the
asymptotic region corresponding to each mode.  It is possible to
obtain the Bogoliubov coefficients from Fourier coefficients
$d_{A,B,C}^{(m)}$ of these modes. We integrate the wave equation
\eqref{eq:wave-eq} from the inner boundary of numerical region
(corresponds to $x=-\infty$) with specified boundary conditions for
$\psi_{A,B,C}$ to the outer numerical boundary (corresponds to
$x=+\infty$) across the sonic horizon $x=0$, and read off coefficients
$d^{(m)}_{A,B,C}$.

\begin{figure}[H] 
  \centering
  \includegraphics[width=1\linewidth,clip]{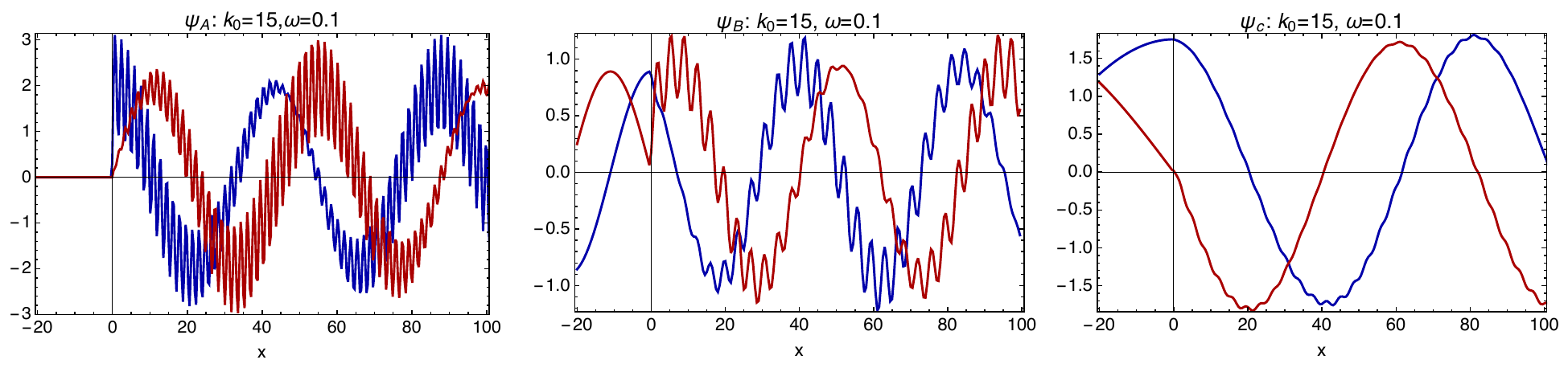}
  \caption{Three mode functions to obtain the Bogoliubov coefficients
    (figure shows $k_0=15,\omega=0.1$ case). Blue lines and red lines
    correspond to the real part and the imaginary part of each wave
    function, respectively. The mode function $\psi_A$ does not show
    oscillatory behavior in $x<0$ region. The mode functions $\psi_B$
    and $\psi_C$ show oscillation in $x<0$ which correspond to
    propagating $\bar u$ mode and $v$ mode, respectively.}
  \label{fig:psi}
\end{figure}

We adopted the 4th order Runge-Kutta method to solve the wave
equation. For numerical calculation, we prepare computation region
$-20\leq x \leq 250$ with discretization $\Delta x=0.001$.  The
parameters of the velocity profile $V(x)$ are $D=0.7, \kappa=7$. The
corresponding Hawking temperature is $T_H=\kappa/(2\pi)=1.11$. The
cutoff parameters in the dispersion are chosen as $k_0=2,15$, which
correspond to $\omega_\text{cutoff}=0.6, 4.5$. Thus these two cutoff
parameter correspond to $\omega_\text{cutoff}<T_H$ for $k_0=2$ and
$\omega_\text{cutoff}>T_H$ for $k_0=15$. Calculated range of the
frequency is $0<\omega<\omega_\text{cutoff}$ with
$\Delta\omega=0.0001$.  Figure \ref{fig:psi} shows behaviors of
$\psi_{A,B,C}$ with $k_0=15, \omega=0.1$; they reflect different
boundary conditions at $x=-20$ which is the inner boundary of our
numerical integration.
\subsection{Result} 

\begin{figure}[H] 
    \centering
    \includegraphics[width=1.0\linewidth,clip]{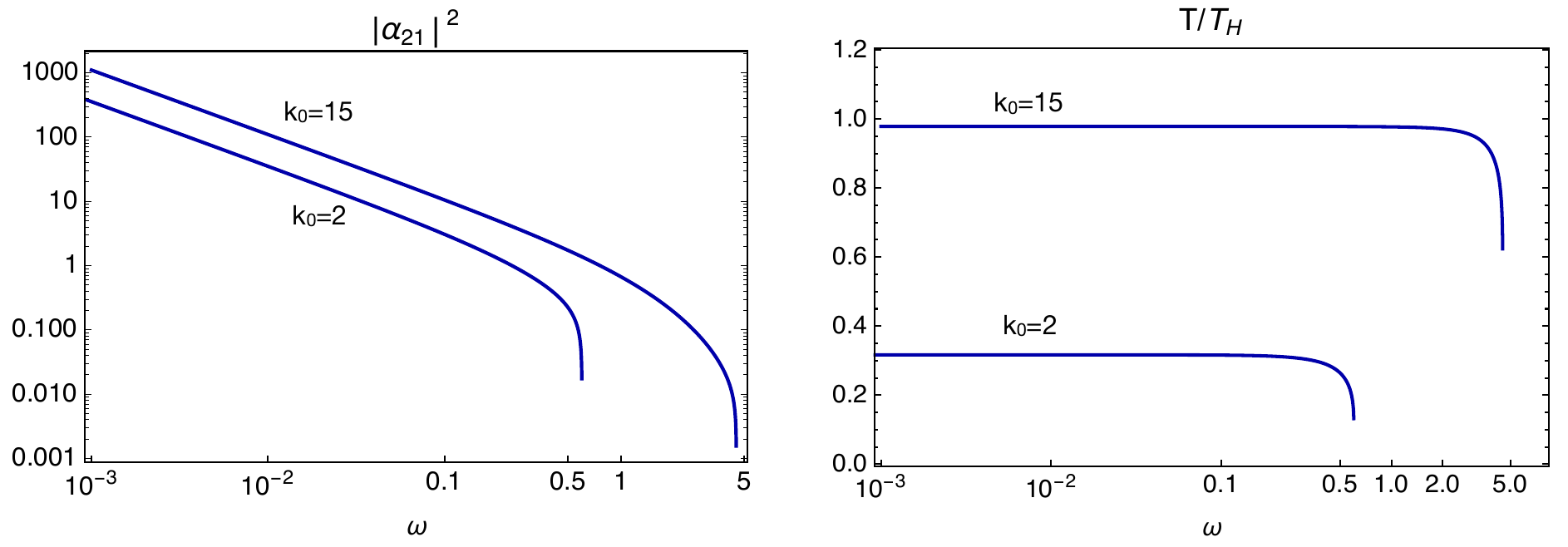}
    \caption{Left panel: the spectrum of created particle number
      density $n_u(\omega)=|\al_{\bu u}|^2$ for $k_0=2,15$. The cutoff frequency is
      $\omega_\text{cutoff}=0.6, 4.5$.  Right panel shows effective
      temperature $T(\omega)$ defined by \eqref{eq:Planck}. For
      $\omega\ll\omega_\text{cutoff}$, the spectrum is well fitted
      with thermal ones but its temperature differs from $T_H$ for
      $k_0=2$. As the cutoff frequency becomes larger, the temperature
      approaches $T_H$. }
    \label{fig:spectrum}
  \end{figure}

\begin{figure}[H]
    \centering
    \includegraphics[width=0.5\linewidth,clip]{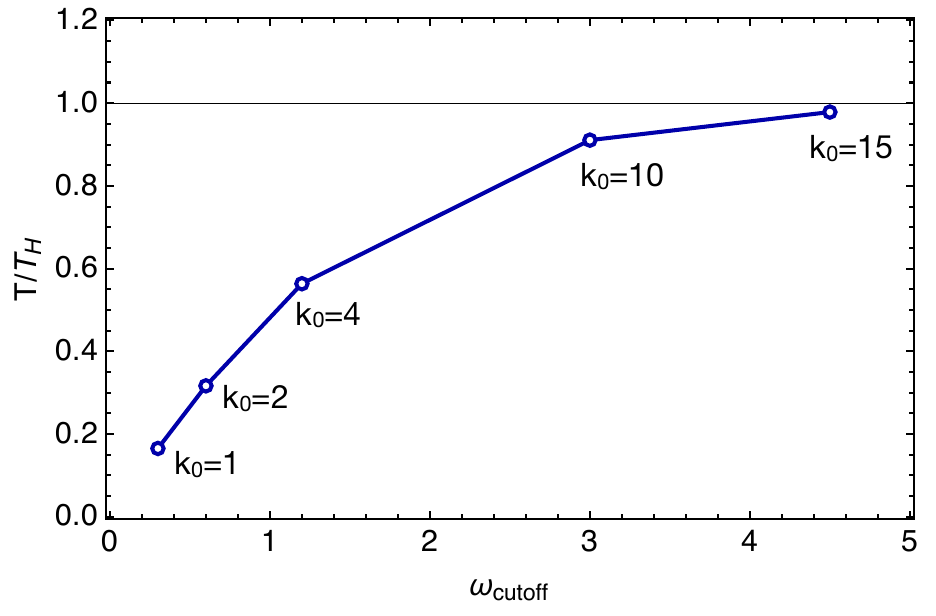}
    \caption{The cutoff parameter dependence of the thermal
      temperature at $\omega=0.001$. Open circles correspond to cutoff
      wave numbers $k_0=1,2,4,10,15$. For
      $\omega_\text{cutoff}<T_H~(k_0<3.71)$, the temperature is
      proportional to $\omega_\text{cutoff}$. For
      $\omega_c>T_H~ (k_0>3.71)$, the temperature asymptotically
      approaches $T_H$ as the cutoff frequency increases.}
    \label{fig:rel} 
  \end{figure}

 From the Bogoliubov transformation \eqref{eq:Bogo-al}, the
  number  of  outgoing Hawking particles (mode $u$) for
   the in-vacuum state $\ket{\psi_0}$ is
   \begin{equation}
    N_u(\omega)=\bra{\psi_0}\hat a_u^\text{out}{}^{\dag}\hat
    a_u^\text{out}\ket{\psi_0}=|\al_{\bu u}|^2\,\del(0)
  \end{equation}
  where the divergent factor $\del(0)$ accounts for an infinite spatial
  volume. The mean number density of particles is
  $n_u(\omega)=|\al_{\bu u}|^2$.  We compare numerically obtained
  $|\al_{\bu u}|^2$ with the following spectrum with $\omega$ dependent
  temperature $T(\omega)$~\cite{Macher2009,Macher2009b}:
  \begin{equation}
    n_u(\omega)=\frac{1}{e^{\omega/T(\omega)}-1}.
    \label{eq:Planck}
  \end{equation}
  Figure \ref{fig:spectrum} shows the spectrum of particle number
  density obtained by our numerical calculation. For
  $\omega\ll\omega_\text{cutoff}$, the spectrum agrees well with the
  thermal one but its temperature differs from the Hawking temperature
  $T_H$ (see Fig.~\ref{fig:rel}). For small cutoff frequency
  $\omega_\text{cutoff}<T_H$, the temperature is proportional to
  $\omega_\text{cutoff}$. As the cutoff frequency becomes larger and
  satisfies $T_H<\omega_\text{cutoff}$, the thermal temperature
  approaches $T_H$~\cite{Macher2009b}. The value of temperature is
  related to the location where the mode conversion occurs. For large
  cutoff $T_H<\omega_\text{cutoff}$, the left moving wave $u_2$ is
  reflected in the vicinity of the sonic horizon and the thermal
  temperature is determined by the ``surface gravity'' of the sonic
  horizon. On the other hand for small cutoff frequency
  $\omega_\text{cutoff}<T_H$, the mode conversion occurs at the
  location \eqref{eq:turning} depart from the sonic horizon and the
  thermal temperature is lowered.

\begin{figure}[H] 
      \centering
      \includegraphics[width=1.\linewidth,clip]{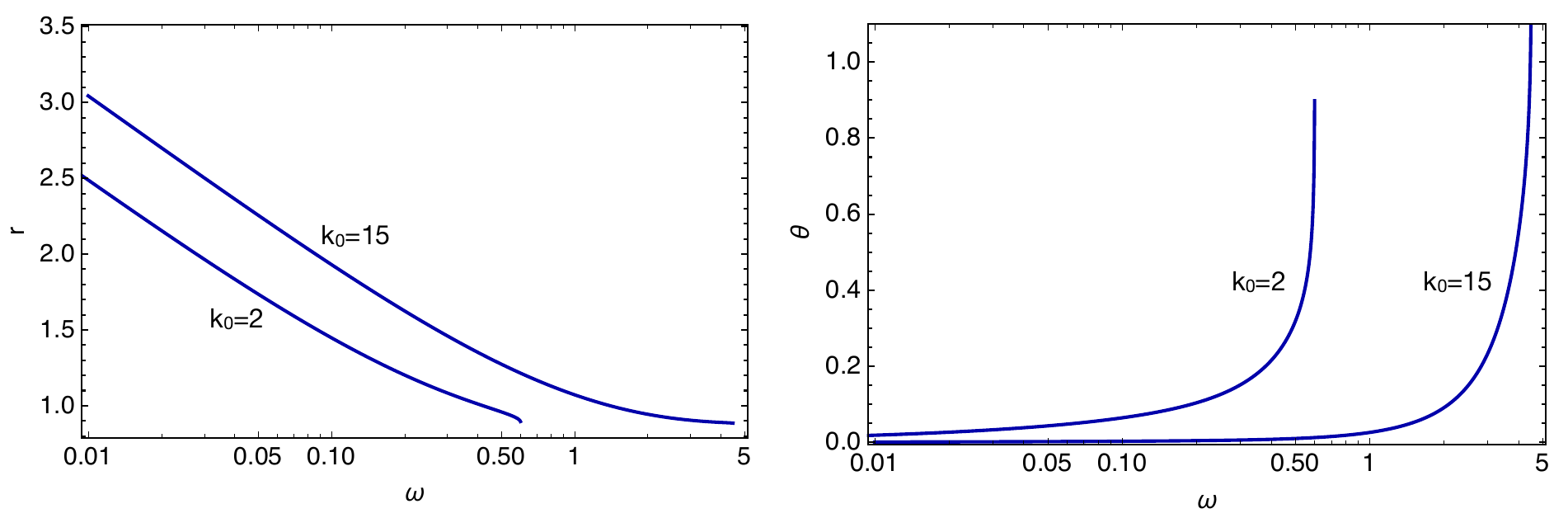}
      \caption{Frequency dependence of parameters $r, \theta$ for
        $k_0=2,15$. Around $\omega\sim \omega_\text{cutoff}$, the
        squeezing parameter $r$ becomes smaller than one and emission of the
        Hawking radiation is shut down.}
      \label{fig:r-theta}
    \end{figure}

  The entanglement structure of the state with the covariance matrix
  \eqref{eq:V0} is encoded in behavior of parameters
  $r(\omega),\theta(\omega)$, which are determined by the Bogoliubov
  coefficients $\al_{\bu u},\al_{\bu\bu},\al_{\bu v}$. Figure \ref{fig:r-theta}
  shows $\omega$-dependence of these parameters obtained from our
  numerical calculation. As the number density of the Hawking
  particles is $|\al_{\bu u}|^2=\cos^2\theta\sinh^2 r$, emission of
  radiation is monotonically decreases as $\omega$ increases.

\begin{figure}[H]
    \centering
    \includegraphics[width=1.0\linewidth,clip]{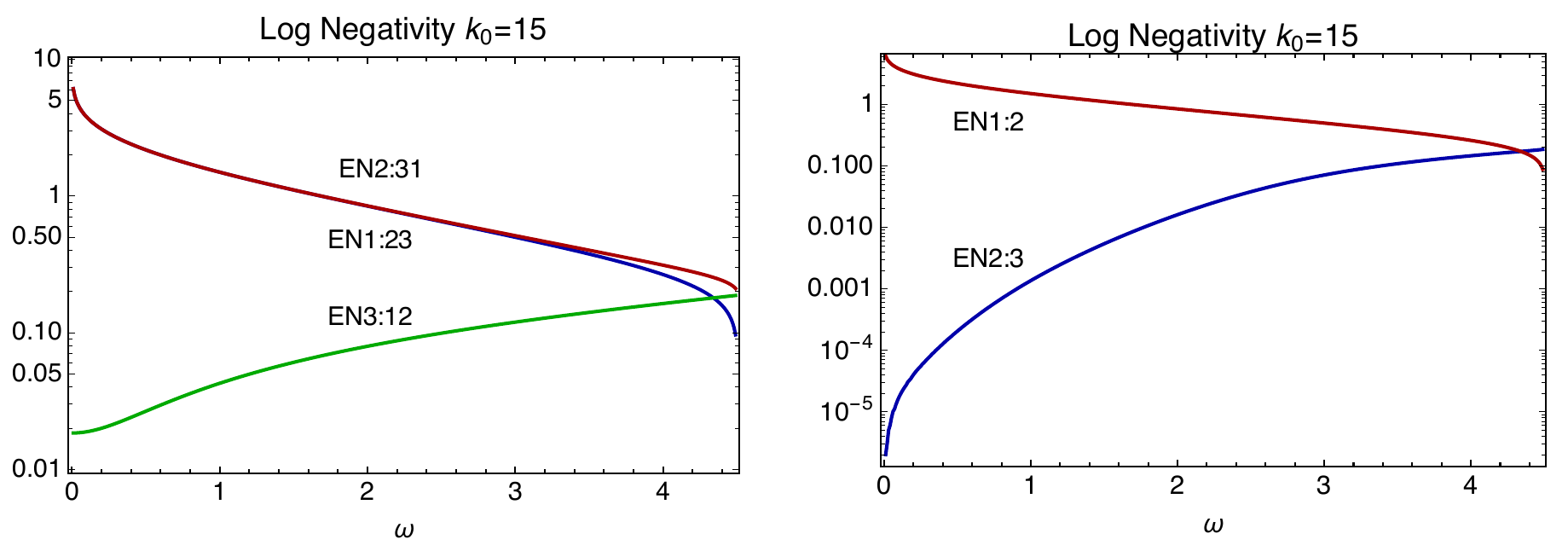}
    \caption{Logarithmic negativity for $k_0=15$
      ($\omega_\text{cutoff}=4.5$). Left panel: logarithmic negativity
      of the three mode state. Right panel: logarithmic negativity of
      reduced two mode states. $E_N{}_{3:1}=0$ holds.}
    \label{fig:lneg15} 
  \end{figure}

\begin{figure}[H]
    \centering
    \includegraphics[width=1.0\linewidth,clip]{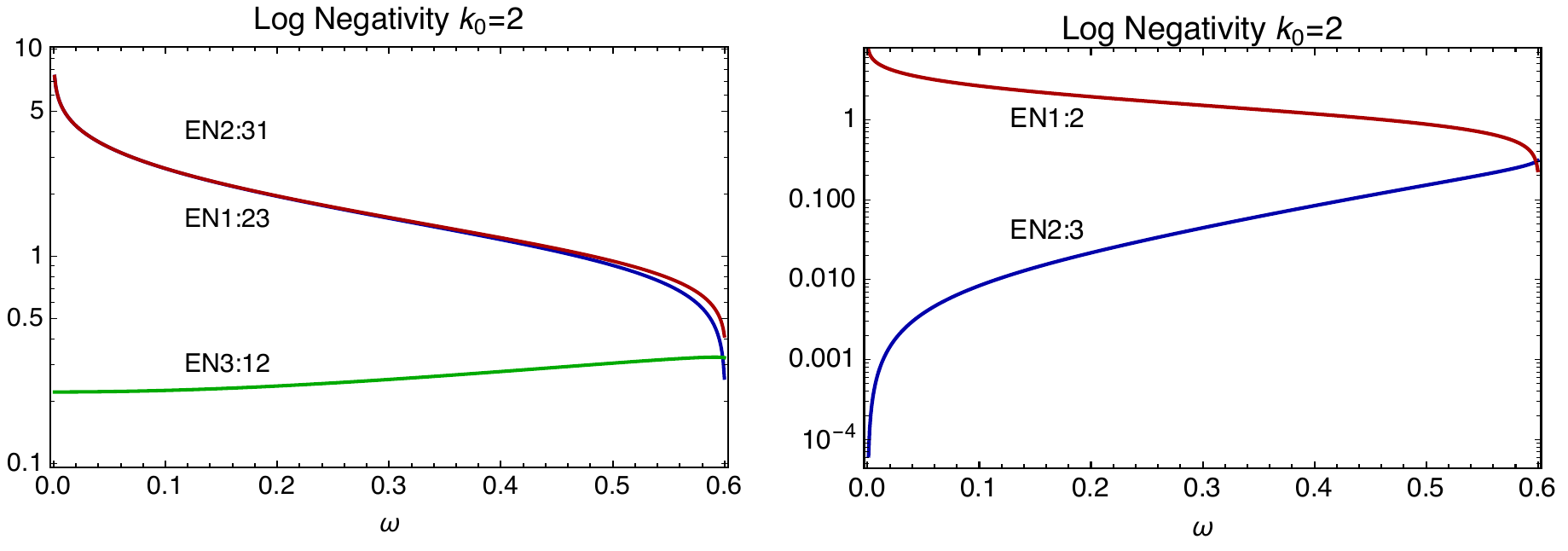}
    \caption{Log negativity for $k_0=2$
      ($\omega_\text{cutoff}=0.6$). Left panel: logarithmic negativity
      of three mode state. Right panel: logarithmic negativity of
      reduced two mode state. $E_N{}_{3:1}=0$ holds.}
    \label{fig:lneg2} 
  \end{figure}

    Figure \ref{fig:lneg15} shows behavior of the logarithmic
    negativity for $k_0=15$ ($T_H<\omega_\text{cutoff}$ case).  Left
    panel shows bipartite entanglement for three different
    bipartitions. For $\omega\ll\omega_\text{cutoff}$,
    $E_N{}_{3:12}\ll E_N{}_{2:31}\approx E_N{}_{1:23}$ and
    entanglement of the system is mainly shared by the mode 1 ($u$) and
    2 ($\bu$). For $\omega\sim\omega_\text{cutoff}$,
    $E_N{}_{1:23}\approx E_N{}_{2:31}\approx E_N{}_{3:12}$ and
    entanglement is equally shared by all three modes. Right panel
    shows bipartite entanglement for reduced two mode states. An
    equality $E_N{}_{3:1}=0$ always holds. We can confirm that the
    mode 1 $(u)$ and 2 $(\bu)$ constitutes a entangled pair for
    $\omega\ll\omega_\text{cutoff}$ because entanglement of the system
    is mainly shared by the mode 1,2 $(u,\bu)$ and the mode 3 $(v)$ decouples.  Figure
    \ref{fig:lneg2} shows behavior of the logarithmic negativity for
    $k_0=2$ ($\omega_\text{cutoff}<T_H$ case). Although values of
    entanglement at $\omega\sim 0$ and
    $\omega\sim \omega_\text{cutoff}$ are ten times larger compared to
    $k_0=15$ case, qualitative dependence of $\omega$ is the same as
    $k_0=15$ case.


\begin{figure}[H] 
      \centering 
      \includegraphics[width=1\linewidth,clip]{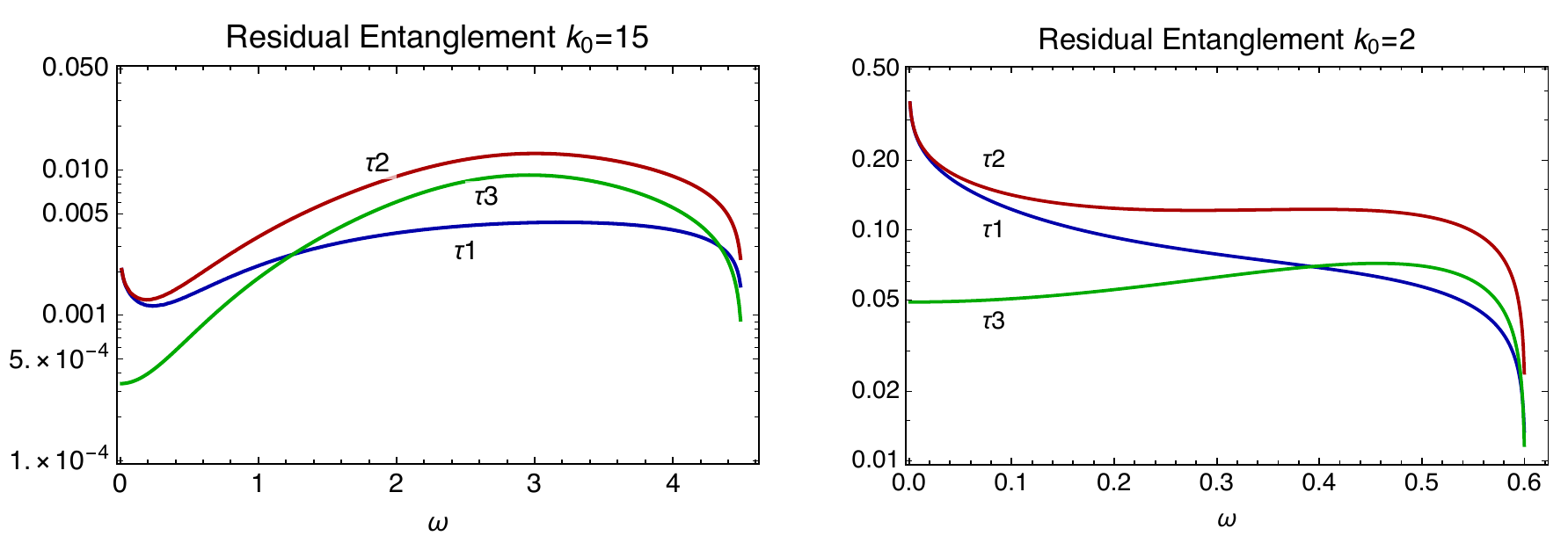}
      \caption{ Behavior of the residual entanglement (left panel:
        $\omega_\text{cutoff}=4.5$, right panel:
        $\omega_\text{cutoff}=0.6$). For
        $\omega<\omega_\text{cutoff}$, the residual $\tau_{1,2,3}$ has
        nonzero value. However, as $\tau_{1,2,3}\ll E_N{}_{3:12}$,
        the tripartite entanglement of the system can be approximately
        reducible to the bipartite entanglement between mode 1 $(u)$ and
        2 $(\bu)$. For $\omega\sim\omega_\text{cutoff}$, $\tau_{1,2,3}$
        approach to nonzero small values and the tripartite
        entanglement of the system can be approximately
        reducible to the bipartite entanglement between mode 1, 2 $(u,\bu)$ and
        2,3 $(\bu,v)$.}
      \label{fig:tau}
    \end{figure}

\begin{figure}[H] 
      \centering 
      \includegraphics[width=0.5\linewidth,clip]{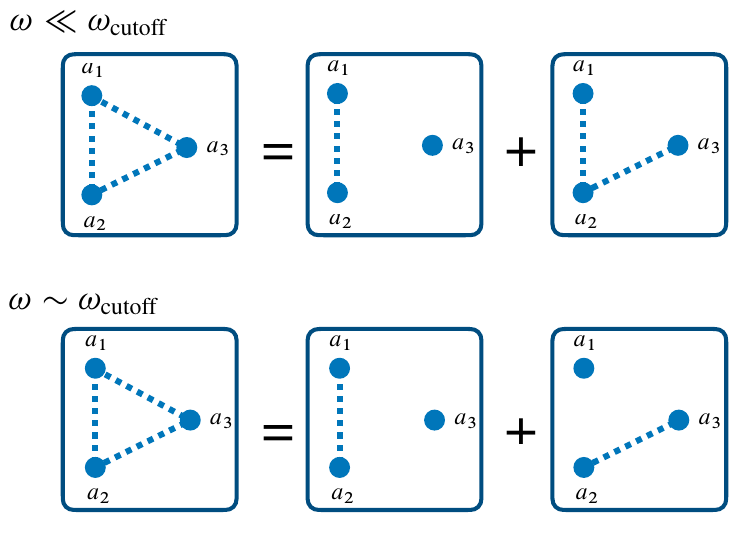}
      \caption{Schematic structure of the tripartite entanglement in
        dispersive model.  For low frequency
        $\omega\ll\omega_\text{cutoff}$, the mode 1 ($u$) and
        2 ($\bu$)  constitute an
        entangled pair because genuine tripartite entanglement is
        small. For $\omega\sim\omega_\text{cutoff}$, genuine
        tripartite approaches zero and entanglement of the system is
        equally shared by the modes 1,2 ($u,\bu$) and the modes
        2,3 ($\bu,v$).}
      \label{fig:tau-sche}
      \end{figure}

  Figure \ref{fig:tau} shows behavior of the residual of
  entanglement. For $\omega<\omega_\text{cutoff}$, the genuine
  tripartite entanglement persists. However, as
  $\tau_{1,2,3}\ll E_{N3:12}$ and $E_N{}_{2:31}\approx E_N{}_{1:23}$,
  $E_N{}_{2:3}<E_N{}_{1:2}$, the entanglement of the system is mainly
  shared by the modes $u$ and $\bar u$. Thus we can conclude that
  $\bar u$ mode is approximately the partner mode of the Hawking
  particle $u$.  For $\omega\sim \omega_\text{cutoff}$, the residual
  of entanglement approaches nonzero small value and the
  entanglement of the system is approximately equally shared by
  pairs of the modes $v, \bar u$ and $u,\bar u$.  Thus the
  tripartite entanglement is reducible to the bipartite entanglement
  between two modes (Fig.~\ref{fig:tau-sche}).
    
      The spectrum of the Hawking radiation around
      $\omega\ll\omega_\text{cutoff}$ well agrees with the thermal one
      with a temperate given by $T\sim T_H$ for
      $\omega_\text{cutoff}>T_H$ and $T\propto \omega_\text{cutoff}$
      for $\omega_\text{cutoff}<T_H$.  From the viewpoint of
      entanglement, although genuine tripartite entanglement exists
      around $\omega<\omega_\text{cutoff}$, its amount is smaller than
      the bipartite entanglement between the mode 1 $(u)$ and 2
      $(\bar u)$. Thus the mode 1 $(u)$ and 2 ($\bar u$) constitute
      entangled pair and the mode 3 $(v)$ decouples.  Thus the
      expected spectrum of emitted radiation (mode $u$) coincides with
      the thermal one because reduced state of the mode $u$ is a state
      of maximal entropy.  For frequency around
      $\omega_\text{cutoff}$, entanglement of the system is equally
      shared by the modes $u,\bu$ and the modes $\bu,v$. Hence
      the reduced state of the mode $u$ does not show the
      thermal feature.

\section{Conclusion}
We examined effect of multipartite entanglement on the Hawking
radiation in the dispersive model and found that the tripartite
entanglement persists in the whole frequency range up to
$\omega_\text{cutoff}$. The origin of this multipartite entanglement
is mixing of the left moving $v$ mode and remaining $u,\bar u$ modes.
Usually, this mixing is recognized as the gray-body factor due
  to back scattering of waves. In the dispersive model investigated in
  this paper, the origin of the gray-body factor is spatially varying
  background flow
  velocity~\cite{Anderson2014,Anderson2015,Fabbri2016}. Thus even for
  low frequency region where the fluid is nearly dispersionless,
  nonzero values of the tripartite entanglement is expected as
  confirmed by our numerical calculation.  The effect of this mixing
and relation to the tripartite entanglement has not been
considered in previous investigations.  For low frequency region, the
mixing between the $v$ mode and $u,\bar u$ modes are small compared to
that between $u$ and $\bar u$, and resulting spectrum of the Hawking
radiation coincides with the thermal one but its temperature depends
on the cutoff scale. Emergence of the thermal feature corresponds to
the structure of entangled pair $u, \bar u$ and decoupling of $v$.
Around the cutoff frequency, the thermal feature of the Hawking
radiation is lost; entanglement of the system is equally shared by
pairs of $v$-$\bar u$ and $u$-$\bar u$. In this region, the in-vacuum
state is a three mode entangled state and this entanglement does not
reduce only to that between the Hawking mode $u$ and its partner
$\bu$.

If we consider the Hawking radiation for gravitational black
holes, $u(\bar u)$-$v$ mixing naturally occurs due to the gray-body
factor which comes from curvature scattering effect. Thus the
  tripartite entanglement of the Hawking radiation is also expected
  but the behavior in low frequency region will depend on type of
  black holes and type of quantum fields. For the massless mininally
  coupled scalar field, the gray-body factor for the $s$-wave in low
  frequency limit is zero for the Schwarzschild black hole and
  nonzero for the Schwarzschild-de Sitter black hole, but the
  behavior is different for the nonminimally coupled scalar
  field~\cite{Kanti2005,Anderson2015}. Thus behavior of the tripartite
  entanglement of the Hawking radiation also depends on the spacetime
  structures and type of quantum fields.

As remaining issues not investigated in this paper, the first one is
effect of different type of dispersion (e.g. superluminal type) on
entanglement. In this paper, we have derived the exact form of the
in-vacuum state involving three modes without specifying the dispersion
relation. As the difference of dispersion is encoded in frequency
dependence of the Bogoliubov coefficients, it is simple task to apply
our formula to other dispersive models to investigate entanglement
structure.  The second one is effect of nonvacuum in-state. If we
assume the thermal state as the in-state, which corresponds to
classical noise from the external environment, the quantum correlation
is reduced and the structure of the multipartite entanglement will be
affected. The third one is the dependence of flow profile $D, \kappa$
on entanglement. The shape of background flow profile may strongly
affects the structure of multipartite entanglement and it is
interesting problem to examine difference of velocity profile on
structure of entanglement sharing of the Hawking radiation.  These
problems are left for our future research.

After uploading our paper to arXiv, we noticed the
  preprint~\cite{Isoard2021}. In this paper, the authors also
  investigate the tripartite entanglement of the Hawking radiation for
  the superluminal dispersion and discuss the best experimental
  configuration for observing entanglement in the BEC system.


\acknowledgements{We would like to thank the anonymous referee for
  helpful comments to improve the paper. 
Y.N. was supported in part by JSPS KAKENHI Grant No. 19K03866. }


\end{document}